\documentclass[usenatbib]{mn2e}

\usepackage{lscape}
\usepackage{graphicx}
\usepackage{amssymb}
\usepackage{amsmath}
\usepackage{url}
\usepackage{aas_macros}

\newcommand{\msun}{\mbox{$\mathrm{M_{\odot}}$}}		
\newcommand{\Uwfc}{\mbox{$U_{_{\mathrm{WFC}}}$}}             
\newcommand{\gwfc}{\mbox{$g_{_{\mathrm{WFC}}}$}}             
\newcommand{\rwfc}{\mbox{$r_{_{\mathrm{WFC}}}$}}             
\newcommand{\iwfc}{\mbox{$i_{_{\mathrm{WFC}}}$}}             
\newcommand{\Zwfc}{\mbox{$Z_{_{\mathrm{WFC}}}$}}             
\newcommand{\giwfc}{\mbox{$(g-i)_{_{\mathrm{WFC}}}$}}       
\newcommand{\riwfc}{\mbox{$(r-i)_{_{\mathrm{WFC}}}$}}       

\usepackage{multirow}

\title[Cluster Collaboration isochrone server]{Pre-main-sequence
  isochrones -- III. The Cluster Collaboration isochrone server}

\author[C.~P.~M.~Bell et al.]{Cameron~P.~M.~Bell,$^{1,2}$\thanks{E-mail:
  cbell@pas.rochester.edu}  Jon~M.~Rees,$^{2}$ Tim Naylor,$^{2}$
N.~J.~Mayne,$^{2}$ R.~D.~Jeffries,$^{3}$
\newauthor Eric~E.~Mamajek$^{1}$ and John~Rowe$^{2}$\\
$^{1}$ Department of Physics \& Astronomy, University of Rochester,
Rochester, NY 14627-0171, USA\\
$^{2}$ School of Physics, University of Exeter, Exeter EX4 4QL, UK\\
$^{3}$ Astrophysics Group, Keele University, Staffordshire ST5 5BG, UK\\}

\begin{document}

\date{Accepted 2014 September 17. Received 2014 September 5; in original form 2014 February 12}

\pagerange{\pageref{firstpage}--\pageref{lastpage}} \pubyear{2014}

\maketitle

\label{firstpage}

\begin{abstract}
We present an isochrone server for semi-empirical pre-main-sequence
model isochrones in the following systems: Johnson-Cousins, Sloan Digital Sky Survey,
Two-Micron All-Sky Survey, Isaac Newton Telescope (INT) Wide-Field
Camera, and INT Photometric H$\alpha$ Survey (IPHAS)/UV-Excess Survey (UVEX).
The server can be
accessed via the Cluster Collaboration webpage
\url{http://www.astro.ex.ac.uk/people/timn/isochrones/}. To achieve this
we have used the observed colours of member stars in young
clusters with well-established age, distance and reddening to create fiducial loci
in the colour-magnitude diagram. These empirical sequences have been
used to quantify the discrepancy between the models and data arising
from uncertainties in both the interior and atmospheric models, resulting
in tables of semi-empirical bolometric corrections (BCs) in the various
photometric systems. The model isochrones made available through the server are based on existing stellar interior models coupled with our newly derived semi-empirical BCs.

As part of this analysis we also present new cluster parameters for both the Pleiades and Praesepe, yielding ages of $135^{+20}_{-11}$ and $665^{+14}_{-7}\,\rm{Myr}$ as well as distances of $132 \pm 2$ and $184 \pm 2\,\rm{pc}$ respectively (statistical uncertainty only).
\end{abstract}

\begin{keywords}
  stars: evolution -- stars: formation -- stars: pre-main-sequence --
  stars: fundamental parameters --
  techniques: photometric -- open clusters and associations: general
  -- Hertzsprung-Russell and colour-magnitude diagrams
\end{keywords}

\section{Introduction}
\label{introduction}

\subsection{Motivation}
\label{intro:motivation}

Our understanding of the physical processes occurring
within stellar interiors has always been driven by the comparison between the
outputs predicted by stellar evolutionary models, e.g. masses,
luminosities and radii, and the observational data. Open clusters thus
provide us with an invaluable testbed with which to assess and constrain
these evolutionary models, based in large part on the assumption that
the stars within such clusters are of a similar age, distance and
chemical composition. Arguably, the most powerful tool for studying clusters is
the colour-magnitude diagram (CMD), which provides a robust test of stellar
physics as it requires the evolutionary models to work consistently
over a broad range of evolutionary states and masses where many
different physical processes are acting concurrently. Furthermore,
CMDs permit the derivation of age, distance and masses for
stellar populations, allowing us to test, for example, whether the initial mass function (IMF) is universal or sensitive to environmental conditions (e.g. \citealp*{Bastian10}).

In a series of papers we have investigated the difficulties associated with using pre-main-sequence (pre-MS) model isochrones to derive properties, such as age and mass, from photometric observations of young stars. \cite{Bell12} (hereafter Paper~I) demonstrated that calibrating photometric observations of pre-MS stars using observations of main-sequence (MS) stars can introduce additional significant and systematic uncertainties. In Paper~I we avoided introducing this additional uncertainty by characterising our adopted natural photometric system. Performing comparisons of the combination of interior and atmospheric models to our observations revealed inherent discrepancies in the optical colours for effective temperatures ($T_{\rm{eff}}) \lesssim 4300\,\rm{K}$. These discrepancies could be as large as a factor of two in the flux at $0.5\,\rm{\mu m}$, resulting in ages derived for young $(<10\,\rm{Myr})$ stars being underestimated by up to a factor of three. Paper~I was just the latest in a series of papers which have demonstrated that at low masses model isochrones fail to match the observed loci of stellar clusters in CMDs (see also \citealp{Baraffe98}, \citealp{Hartmann03}, \citealp{Vandenberg03}, \citealp*{Bonatto04}, \citealp{Pinsonneault04} and \citealp{Stauffer07}). This is also demonstrated in Fig.~\ref{fig:pleiades_cmd} (see Section~\ref{comparison_with_literature_relations}) where the dashed line representing a theoretical isochrone is too blue when compared with photometry of Pleiades members.

Having quantified the discrepancy between the models and the data as a function of $T_{\rm{eff}}$ in individual photometric bandpasses, \cite{Bell13} (hereafter Paper~II) demonstrated that adopting these empirical corrections resulted in pre-MS model isochrones that matched the observed shape of the pre-MS locus. Using a sample of 13 young clusters, the age of each was assessed from two sets of model isochrones. First those that follow the nuclear evolution of the relatively massive MS stars, and second those that follow the contraction of low-mass stars to the MS. Adopting our new semi-empirical pre-MS model isochrones, we obtained, for the first time for a set of clusters, consistency between these two age diagnostics for clusters younger than $30\,\rm{Myr}$ (see also the studies of \citealp{Lyra06} and \citealp*{Pecaut12}).

The recent study of NGC\,1960 by \cite{Jeffries13} demonstrates that we are now beginning to establish a reliable age scale for young clusters with ages above $20\,\rm{Myr}$ from measurements of the lithium depletion boundary (LDB; see also \citealp{Soderblom13}), with consistency shown between these ages and those from the MS and pre-MS members of the same cluster. Whilst there is a lower limit to the applicability of the LDB ageing method ($\simeq 20\,\rm{Myr}$) due to a higher level of model dependency below this, Paper~II highlighted continued consistency between ages derived from MS and pre-MS isochrones down to $\simeq 10\,\rm{Myr}$. Thus, we feel the time is now right to make the semi-empirical pre-MS model isochrones we have used to establish this consistency publicly available through an isochrone server.

\subsection{Methodology}
\label{intro:methodology}

\begin{figure}
\centering
\includegraphics[width=\columnwidth]{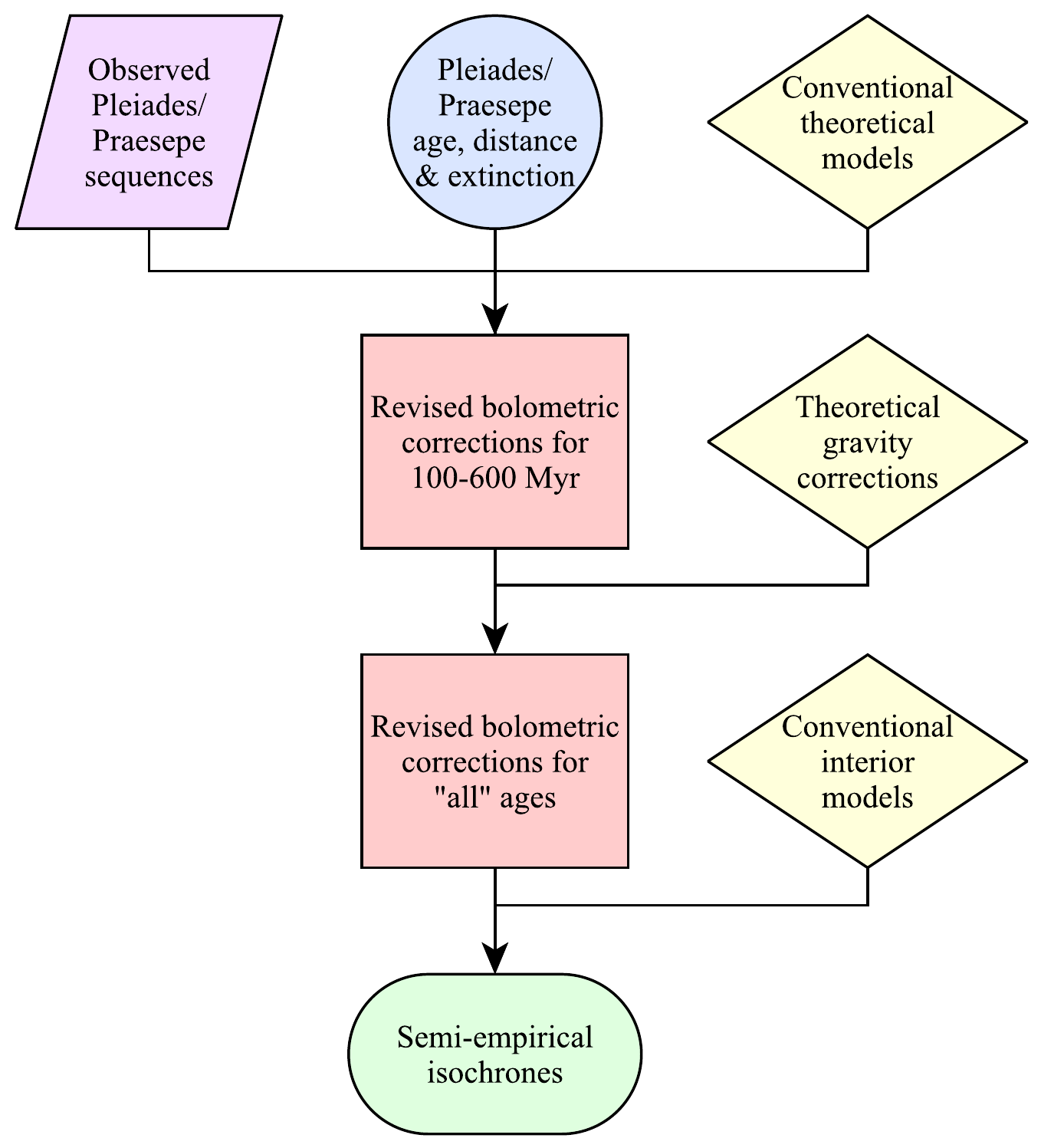}
\caption[]{Flowchart summarising the main steps in creating the semi-empirical pre-MS model isochrones. The purple parallelogram represents the fiducial Pleiades and Praesepe loci, created from observed colours of member stars in each cluster (see Section~\ref{the_data}). The blue circle denotes the input age, distance and extinction for the Pleiades and Praesepe (see Section~\ref{fiducial_loci_parameters}). The yellow diamonds signify the theoretical models (both interior and atmospheric; see Section~\ref{the_models}). The red rectangles represent the processes of creating the semi-empirical BCs in various photometric systems by comparing the model isochrones to the fiducial loci in CMD space (see Section~\ref{creating_semi-empirical_pre-MS_isochrones}). Finally, the green oval denotes the semi-empirical pre-MS model isochrones which we make available through the server (see Section~\ref{internet_server}).}
\label{fig:flowchat}
\end{figure}

The use of empirical isochrones to better constrain stellar ages and distances is not unprecedented, and has been applied across a wide range of evolutionary phases including pre-MS (e.g. \citealp{Stauffer98a}; \citealp*{Jeffries01}), older MS (e.g. \citealp{Pinsonneault04,An07}) and globular clusters (e.g. \citealp{Brown03,Brown04}).
There are several steps involved in creating the semi-empirical pre-MS model isochrones presented in the server, and although these have been detailed in Papers~I and II, we feel it is best to reiterate these steps below, in addition to the underlying assumptions which these models are based upon. These steps are summarised in Fig.~\ref{fig:flowchat}.

In Section~\ref{the_data} we present the photometric systems for which we can create a fiducial locus in CMD space and discuss possible systematic effects on the photometry arising from transformations between different systems. The semi-empirical models rely on us calculating the observed discrepancy between the data and the theoretical model isochrones, and therefore in Section~\ref{the_models} we describe the interior and atmospheric models we have used. In addition to the theoretical model isochrones, we also require parameters for our fiducial clusters (age, distance and reddening) and these are discussed in Section~\ref{fiducial_loci_parameters}.

To calculate quantitatively the mismatch between the data and the models, we need to measure this discrepancy in individual bandpasses. In Paper~I we used a sample of low-mass binaries with well-constrained dynamical masses and multi-band photometry to test whether the combination of interior and atmospheric models could reliably predict the combined system magnitudes. This showed that whilst the models over-predict the optical fluxes, the predictions for the $K_{\rm{s}}$-band were essentially correct. We therefore created our colours with respect to the $K_{\rm{s}}$-band. Hence, assuming that the models reliably predict the relationship between the $K_{\rm{s}}$-band absolute magnitude and the mass (which appears a reasonable assumption in the mass range covered by our fiducial clusters; see e.g. \citealp{Delfosse00}) then this ensures that the resultant semi-empirical model isochrones have a reliable mass scale which is tied to that of the fiducial cluster. \emph{Assumption 1: the theoretical models fit the $K_{\rm{s}}$-band flux well and therefore we can use this to determine the $T_{\rm{eff}}$ at points along our fiducial locus and derive the necessary corrections}.

Our fiducial locus determines the empirical corrections (here termed $\Delta$BCs) to the theoretical models as a function of $T_{\rm{eff}}$ for a specific age (or equivalently surface gravity; log$\,g$). Thus, if we are to provide semi-empirical model isochrones at different ages we must assume that the cause of this discrepancy does not depend on log$\,g$. \emph{Assumption 2: the empirical corrections in a given bandpass have the same log$\,g$ dependence as the theoretical bolometric corrections (BCs).} A further assumption concerning the creation of the semi-empirical model isochrones and their application to clusters spanning a wide age range relates to intrinsic differences between clusters of different ages. \emph{Assumption 3: all clusters are assumed to behave like the Pleiades such that there are no intracluster effects due to, for example, binary fractions and mass ratios, intrinsic age spread, rotation distribution, and activity-related effects on the observed colours}.

Section~\ref{creating_semi-empirical_pre-MS_isochrones} describes the process of quantifying this discrepancy between the uncorrected models and the data in more detail. For reasons discussed in Paper~I we choose the Pleiades as our default fiducial cluster for all photometric systems discussed here. For the Sloan Digital Sky Survey (SDSS) system, however, there are no Pleiades data and so in Section~\ref{praesepe_suitability} we discuss what effect adopting a different cluster (Praesepe) has on our semi-empirical model isochrones.

Our isochrone server is introduced in Section~\ref{internet_server} and we give a brief overview of its use. Our primary aim in making these model isochrones available is to allow others to measure cluster ages from the pre-MS which are consistent with those we have derived. In addition, these models can also be used to identify possible new cluster members from their positions in CMDs which would then require spectroscopic follow-up. Finally, these models could also be used to calculate mass functions for stellar populations based on their positions in the CMD. We advise users that given the processes involved in creating these model isochrones, they should \emph{not} be used to compare against new cluster photometry and then argue that, due to a mismatch between the models and the data, there is a particular problem with either the interior or atmospheric models.

\section{The Data}
\label{the_data}

In this Section we discuss the photometric systems for which we are
able to construct a well-populated fiducial locus in CMD space using data
from a young cluster with known age and distance. Here, we only discuss the data and memberships adopted, whilst in Section~\ref{fiducial_loci_parameters} we discuss the adopted fiducial cluster parameters and in Section~\ref{semi-empirical_bolometric_corrections} we describe the process of creating the single-star locus.
In this paper we will focus on five photometric systems: Johnson-Cousins, Isaac Newton Telescope Wide-Field Camera (INT-WFC), INT Photometric H$\alpha$ Survey (IPHAS)/UV-Excess Survey (UVEX), Two-Micron All-Sky Survey (2MASS), and Sloan Digital Sky Survey (SDSS).

\subsection{Johnson-Cousins}
\label{data:johnson_cousins}

Our starting point for a Pleiades catalogue in the Johnson-Cousins
photometric system is the $BVI_{\rm{c}}$ catalogue of \cite{Stauffer07}. The
membership catalogue of \citeauthor{Stauffer07} represents a number of
photometric, photoelectric and photographic surveys which have been
combined with a series of proper motion surveys to assign cluster membership.
The reader is directed to Appendix~A1 of
\cite{Stauffer07} for a full and comprehensive discussion of the
catalogues used.

To include $R$-band photometry we adopted
Johnson $(R-I)_{_{\rm{J}}}$ measurements for stars brighter than $V \simeq 11$ from
\cite{Mendoza67}. These Johnson colour indices were converted to
Cousins indices using the following relation from \cite{Bessell79}:

\begin{equation}
(R-I)_{\mathrm{c}}=0.856(R-I)_{_{\mathrm{J}}}+0.025+\Delta(R-I)_{_{\mathrm{J}}},
\end{equation}

\noindent where $\Delta(R-I)_{_{\rm{J}}}$ is a non-linear correction
curve and for which the transformation is accurate to within $0.02\,
\rm{mag}$ for MS stars with colours $(R-I)_{\rm{c}} \lesssim 0.7$. For stars fainter than $V \simeq
11$ we used the photometric data from
\cite{Stauffer80,Stauffer82,Stauffer84} and \cite{Stauffer87}. These
studies present $(RI)_{_{\rm{K}}}$ photometry in the Kron system and so these too
were transformed into Cousins colour indices using a relation from
\cite{Bessell87}:

\begin{eqnarray}
\label{eqn:bessell}
(R-I)_{\mathrm{c}} & = & 0.102+0.9166(R-I)_{_{\mathrm{K}}} \\
& + & 0.4230(R-I)_{_{\mathrm{K}}}^{2}-0.16647(R-I)_{_{\mathrm{K}}}^{3}, \nonumber
\end{eqnarray}

\noindent which is accurate to within $0.02\,\rm{mag}$ for MS stars with colours $(R-I)_{\rm{c}} \lesssim 2.2$. Additional Cousins $(RI)_{\rm{c}}$ photometry for very
low-mass stars and brown dwarfs was taken from \cite{Bouvier98},
however we only retained confirmed proper motion members as defined in
\cite*{Moraux01}.

At the age of the Pleiades the majority of stars have reached
the MS, however those with masses below $\sim 0.4\,\msun$ are still undergoing pre-MS contraction. In Paper~I we discussed the problems with using MS relations to transform photometric observations of pre-MS stars which stem from a combination of differences in log$\,g$ as well as in the spectra between MS and pre-MS stars of the same colour. Hence, what effect does using MS relations to transform the Kron $(RI)_{_{\rm{K}}}$ photometry of the low-mass pre-MS members have on our fiducial sequence in the Cousins $(RI)_{\rm{c}}$ colours?

Using the \cite{Dotter08} and updated Pisa evolutionary models (see Section~\ref{interior_models}), we estimate that the difference in log$\,g$ between a $0.1\,\msun$ star (approximately the lowest mass in the Kron photometric sample) at an age of $130\,\rm{Myr}$ and its arrival on the ZAMS ($\sim 20\,\rm{Gyr}$; \citealp*{Siess00}) is $\sim 0.2-0.3\,\rm{dex}$. Assuming this difference in log$\,g$, we can roughly quantify the possible systematic uncertainty introduced into the transformed photometry using the derived $(RI)_{\rm{c}}$ bolometric corrections (see Section~\ref{atmospheric_models}). From these, we estimate that the uncertainty in the $(R-I)_{\rm{c}}$ colour as a result of the differences in the log$\,g$ between a pre-MS and ZAMS star of mass $\simeq 0.1\,\msun$ is $\le 0.05\,\rm{mag}$ (allowing for slight variations in $T_{\rm{eff}}$ depending on which evolutionary model is adopted). Contrast this with a difference of $\simeq 0.1-0.15\,\rm{mag}$ at an age of $3\,\rm{Myr}$.

Such a systematic uncertainty is not ideal, however given that our choice of populated fiducial clusters is so limited, coupled with the fact that the Cousins $(RI)_{\rm{c}}$ bandpasses have surpassed the Kron $(RI)_{_{\rm{K}}}$ bandpasses as the de facto choice of optical $RI$ bandpasses, there is little option but to proceed with the transformed photometry. We note, however, that we are correcting for an observed discrepancy of $\simeq 0.3\,\rm{mag}$, and that after having done so, the recalibrated model isochrones have a residual systematic uncertainty at the $0.05\,\rm{mag}$ level in $(R-I)_{\rm{c}}$. Furthermore, it is worth noting that the $V-I_{\rm{c}}$ colours given in the \cite{Stauffer07} catalogue represent values transformed using a relation from \cite{Bessell87} based on the same MS sample as those used to derive Eqn.~\ref{eqn:bessell}, however no discussion of introduced systematic uncertainties is given in the paper.

\subsection{Isaac Newton Telescope Wide-Field Camera}
\label{data:int_wfc}

The Pleiades data in the natural photometric system of the INT-WFC is that described in
Papers~I and II. As the photometric observations presented in
these studies extend to lower masses than the Johnson-Cousins
sequence it was necessary to supplement
the \cite{Stauffer07} membership catalogue with lower mass members
from \cite*{Lodieu12} (see Paper~I for a discussion).

A large benefit of including the INT-WFC in this study is due to the WFC's use in both the INT Photometric H$\alpha$ Survey (IPHAS; \citealp{Drew05}) and UV-Excess Survey (UVEX; \citealp{Groot09}). These combined surveys use a combination of the $(Ugri)_{_{\rm{WFC}}}$ and narrow-band $H\alpha$ bandpasses to survey a sky area of approximately 1800 square degrees in the northern Galactic plane spanning the latitude range $-5^{\circ} < b < +5^{\circ}$ to a limit $\rwfc \simeq 20$. Unfortunately, we do not have useful $\Uwfc$-band photometry nor $H\alpha$ observations of the Pleiades and hence these bandpasses are not included in the following discussion.

Whilst our own observations are calibrated to an AB magnitude system to replicate the SDSS (at least approximately), the IPHAS and UVEX surveys are calibrated onto a Vega system. To distinguish between these two systems we will refer to the IPHAS/UVEX system in addition to our own INT-WFC system. The difference between the two is a simple zero-point offset, which is given in Table~\ref{tab:int-wfc_iphas-uvex}, calculated by folding the reference spectrum of Vega and the AB zero-point flux distributions (see Section~\ref{atmospheric_models}) through the system responses given in Paper~I. The differences between these and the values given in Table~1 of \cite{Gonzalez-Solares08} are likely attributable to small differences in the adopted responses.

\begin{table}
\caption[]{The bandpass specific zero-point shifts required to convert magnitudes in the INT-WFC (AB) photometric system to the IPHAS/UVEX (Vega) system.}
\centering
\begin{tabular}{c c}
\hline
Bandpass&INT-WFC -- IPHAS/UVEX\\
\hline
$\gwfc$&-0.090\\
$\rwfc$&+0.147\\
$\iwfc$&+0.394\\
\hline
\end{tabular}
\label{tab:int-wfc_iphas-uvex}
\end{table}

\subsection{Two-Micron All-Sky Survey}
\label{data:2mass}

The 2MASS \citep{Skrutskie06} data
cover an area of $2^{\circ} \times 3^{\circ}$ centred on the Pleiades
cluster and represent a subset of the so-called ``$6\times$''
observations\footnote[1]{A general overview as well as
additional information concerning the ``$6\times$" observations, data
reduction, and catalogue can be found at
\url{http://www.ipac.caltech.edu/2mass/releases/allsky/doc/seca3_1.html}}
(see \citealp{Cutri12}). These observations were taken towards the end of the 2MASS survey with exposures 6 times longer than those used for the primary survey. We have adopted the members
defined in \cite{Stauffer07}, for which memberships have been assigned
in an identical fashion to those described in
Section~\ref{data:johnson_cousins}. 

\subsection{Sloan Digital Sky Survey}
\label{data:sdss}

The Pleiades cluster was not observed as part of the SDSS \citep{York00} and therefore we require an alternative cluster. 
Cross-correlating the positions of the \cite{Dias02} open cluster
catalogue with the Ninth Data Release (DR9) of the SDSS \citep{Ahn12}, we find only two other suitable cluster candidates, namely IC\,4665 and Praesepe.
Unfortunately, although IC\,4665 has an LDB age,
the sequence in CMD space (defined using the \citealp{Jeffries09b} members) is
extremely sparse and does not provide a well-sampled population of stars
across a significant colour range. Praesepe, therefore, offers an appealing alternative on the basis of the following: i) it is a rich cluster, ii) it is nearby, thereby allowing us to access the lowest mass members, iii) it has a very low uniform reddening, and iv) it has a well-defined sequence in the CMD.

We used the Praesepe members as defined by \cite{Kraus07} with
membership probabilities of $\geq 95$ per cent. Using this refined membership list we selected SDSS
photometry for each source by requiring that: i) the object is
classified as a star (``Type''$=6$), ii) the uncertainty on the
magnitude in each bandpass is $\leq 0.1\,\rm{mag}$, and iii) the
photometry flags ``BLENDED'' (whether the object is a composite), ``EDGE'' (whether the object was too close to the edge of the frame) and ``SATURATED'' (whether the object contains any saturated pixels) are all false in each bandpass.

Note that there are two specific issues which need to be addressed when adopting Praesepe as our alternative fiducial cluster. First there is the issue of increased magnetic activity at younger ages (see e.g. \citealp{Stauffer03}) and what effect this may have on the observed colours of stars (in relation to those in the Pleiades). Secondly there is the matter of metallicity and the possible systematic residuals introduced by using a non-solar composition locus to recalibrate solar metallicity pre-MS model isochrones. In Section~\ref{praesepe_suitability} we provide a detailed discussion of both of these issues and demonstrate that, when compared to the observed discrepancy between the models and the data, these effects are not significant.

\section{The Models}
\label{the_models}

Before we can quantify the discrepancy between the theoretical model isochrones and the observed colours of the fiducial loci in CMD space, we must first create the model isochrones using a combination of interior and atmospheric models. The atmospheric models are used to convert the bolometric luminosities ($L_{\rm{bol}}$), $T_{\rm{eff}}$
and log$\,g$ predicted by the
interior models into observable colours and magnitudes. This
conversion is based on BC-$T_{\rm{eff}}$ relations which can be derived
by folding the flux distribution of theoretical atmospheric models through the
appropriate photometric system responses. Here we describe the stellar interior and atmospheric models used in this paper.

\subsection{Stellar interior models}
\label{interior_models}

In addition to the pre-MS model isochrones which we make available via the isochrone server, we also require MS models to derive the cluster parameters for Praesepe and ensure that these are consistent with those we have adopted for the Pleiades. To ensure consistency with the MS ages we have derived in Paper~II, we adopt the same resampled grid of \cite{Lejeune01}, specifically the basic `c' grid \citep{Schaller92}, with spacing $\Delta\rm{log(age)=0.02}$. In the pre-MS regime, we used the
\cite{Baraffe98} with a solar-calibrated mixing-length parameter
$\alpha=1.9$ and the \cite{Dotter08} models (hereafter BCAH98
$\alpha=1.9$ and DCJ08 respectively).

The semi-empirical pre-MS model isochrones created in this paper and distributed via the server comprise of semi-empirical BCs coupled with existing pre-MS interior models. The interior models we choose to adopt for the server consist of the BCAH98 $\alpha=1.9$, DCJ08 and Pisa groups \citep*{Tognelli11}. In Paper~I we were unable to test these models
as they did not extend to the age of the Pleiades, however Emanuele
Tognelli has kindly computed an updated set of Pisa models
spanning a mass range $0.08-9\,\msun$ and extending to ages of
$\sim10\,\rm{Gyr}$ at $1\,\msun$ (priv. comm.). To ensure that the updated Pisa
models predict ages that are consistent with the ages derived from the higher mass MS stars, we have used the $\tau^{2}$ fitting
statistic (see \citealp{Naylor06}) to calculate pre-MS ages for the five young clusters we previously fitted
in Paper~II; namely $\lambda$~Ori, NGC~2169, NGC~2362,
NGC~7160, and NGC~1960. We find that the ages derived using the
updated Pisa models agree with those calculated using the DCJ08 models
to within $\pm 1\,\rm{Myr}$, and are therefore generally consistent with the MS
ages for the regions as calculated in Paper~II.

The basic input assumptions concerning the stellar interior structure and physics
are the same as those described in \cite{Tognelli11}, however we describe some recent updates. The standard Pisa models adopted the equation of state
(EOS) of \cite{Rogers02} to the limiting mass of $0.2\,\msun$. The
updated models, however, extend to much lower masses and so
for masses below $0.2\,\msun$ they are computed using the EOS of
\cite*{Saumon95}.
For the outer boundary conditions, the BT-Settl
atmospheric models of \cite*{Allard11} were used for $T_{\rm{eff}} <
2000\,\rm{K}$, the \textsc{phoenix}/\emph{GAIA} models of
\cite{Brott05} for $2000 \leq T_{\rm{eff}} < 10\,000\,\rm{K}$, and the
\textsc{atlas9}/ODFnew models of \cite{Castelli04} for $T_{\rm{eff}}
\geq 10\,000\,\rm{K}$. The models use the OPAL EOS 2005 radiative
opacity table (see e.g. \citealp{Iglesias96}) for log $T_{\rm{eff}} \geq 4.5$,
and for cooler temperatures that of \cite{Ferguson05}. Note
that both opacity tables have been computed using the more recent
\cite{Asplund09} heavy element solar abundances. The mixing length
parameter $\alpha$ has been tuned to fit the observed properties of
the present day Sun and is calculated to be $\alpha=1.74$. The updated interior
models also include the effects of convective core overshooting --
the quantity which characterises the distance that
convective elements penetrate beyond the classical core in terms of
pressure scale heights --  which has been set to $\lambda_{\rm{over}}=0.2$ for
masses greater than $1.1\,\msun$ (see \citealp*{Tognelli12}).

\subsection{Atmospheric models}
\label{atmospheric_models}

The atmospheric models used are the same as those in
Papers~I and II. For $T_{\rm{eff}} < 8000\,\rm{K}$ we adopt the \textsc{phoenix} BT-Settl models \citep{Allard11}, whereas for $8000 \leq T_{\rm{eff}} \leq 50\,000\,\rm{K}$ we use the Kurucz \textsc{atlas9}/ODFnew models \citep{Castelli04}.

\begin{table}
\caption[]{Photometric systems (and the individual bandpasses)
  adopted in this paper.
  The reference directs the reader to the source for the system
  responses adopted for each system.}
\centering
\begin{tabular}{c c c}
\hline
Photometric&System&Reference\\
system&responses&\\
\hline
Johnson-Cousins&$BV(RI)_{\rm{c}}$&\cite{Bessell12}\\
INT-WFC&$(griZ)_{_{\rm{WFC}}}$&Paper~I\\
SDSS&$griz$&\cite{Doi10}\\
2MASS&$JHK_{\rm{s}}$&\cite{Cohen03}\\
\hline
\end{tabular}
\label{tab:photometric_bandpasses}
\end{table}

We derive BCs
for the Johnson-Cousins, INT-WFC, 2MASS, and SDSS photometric
systems using Eqn.~B2 in Appendix~B of
Paper~I for which we adopt the system responses of
\cite{Bessell12}, Paper~I, \cite*{Cohen03}, and \cite{Doi10} respectively
(see Table~\ref{tab:photometric_bandpasses}).
To calculate BCs for each
photometric system we require a reference
spectrum which produces a known magnitude in a given bandpass. For the
Johnson-Cousins BCs we used the CALSPEC Vega reference spectrum
alpha\_lyr\_stis\_005\footnote[2]{\url{http://www.stsci.edu/hst/observatory/cdbs/calspec.html}}
with $V=0.03$ and all colours equal to zero. In the 2MASS photometric
system we used the bandpass specific reference spectra
and the zero-point offsets of \cite{Cohen03}. Finally, for the INT-WFC
and SDSS systems we used a reference spectrum of constant flux
density per unit frequency $f^{\circ}_{\nu}$ (see e.g. \citealp{Oke83})
converted into energy per unit wavelength according to
$f_{\lambda}=c/\lambda^{2} f_{\nu}$ which results in reference
magnitudes of zero in all bandpasses. In addition to the reference spectra, we
also require solar values for the absolute bolometric magnitude and luminosity
for which we adopt $M_{\rm{bol}, \odot}=4.755\,\rm{mag}$ and $L_{\odot}=3.827 \times 10^{33}\,\rm{erg\,s^{-1}}$ as given in \cite{Mamajek12}.

\section{The age and distance of our fiducial clusters}
\label{fiducial_loci_parameters}

To use the empirical colours of the fiducial loci in CMD space to quantify the discrepancy between the theoretical models and the observed data, we must first adopt a set of parameters (age, distance and reddening) representative of the cluster and apply these to the theoretical model isochrones.

\subsection{The Pleiades}
\label{pleiades_parameters}

 In the case of the Pleiades we
adopt (as in Paper~I) a distance modulus $dm=5.63\,\rm{mag}$ derived
via space-based trigonometric parallax measurements
\citep{Soderblom05}, an age of $130\,\rm{Myr}$ based on detecting the
LDB in the $K_{\rm{s}}$-band
\citep*{Barrado04b}, and a reddening of $E(B-V)=0.04\,\rm{mag}$ \citep{Stauffer98a}\footnote[3]{We independently estimate the reddening to the Pleiades through two methods. First, using the Q-method of \cite{Johnson53} with the homogenised photometry of 15 B-type Pleiades members from \cite{Mermilliod06}, in conjunction with the revised Q/$U-B$/$B-V$ sequences of \cite{Pecaut13}, we estimate a mean reddening of $E(B-V)=0.036 \pm 0.006\,\rm{mag}$. Secondly, using the $ubvy$ photometry of 15 B-type Pleiades members from \cite{Hauck98} and de-reddening following \cite{Castelli91}, and the $E(b-y)$/$E(B-V)$ slope from \cite*{Taylor08}, we estimate a mean reddening of $E(B-V)=0.044 \pm 0.005\,\rm{mag}$. Both of these estimates are in agreement with the value of $E(B-V)=0.04\,\rm{mag}$ we have adopted from \cite{Stauffer98a}.}.


\subsection{Praesepe}
\label{praesepe_parameters}

Praesepe is a much more evolved cluster and therefore we are unable to make use of the LDB to calculate an age. In addition, although the revised \textit{Hipparcos} distance modulus of $dm=6.30\,\rm{mag}$ \citep{VanLeeuwen09} is in good agreement with other determinations (e.g. \citealp{An07}), the uncertainty in this distance modulus (almost $0.1\,\rm{mag}$) is too large for the purposes of calculating empirical BCs at low $T_{\rm{eff}}$. Hence, we derive a consistent age and distance for Praesepe from fitting photometry of MS members in the $V, B-V$ CMD using MS evolutionary models.

Age and distance estimates will be affected by both reddening and metallicity. As Praesepe is significantly older than the Pleiades the higher mass members have already evolved off the MS and entered the post-MS evolutionary phase. Hence, we are unable to derive a reddening using these members in the 
$U-B, B-V$ colour-colour diagram (see e.g. \citealp{Mayne08}). Instead we adopt the mean value of $E(B-V)=0.027\,\rm{mag}$ derived using
a combination of polarisation measurements, comparison of $\beta$ and
$(R-I)_{\rm{c}}$ for F stars, and Str{\"o}mgren $\beta$ analysis of A stars
\citep{Taylor06}.

An additional consideration is the chemical composition of Praesepe. 
Relative to the composition of the Pleiades, which to within the uncertainties is solar ($\mathrm{[Fe/H]}=+0.03 \pm 0.02 \pm 0.05\,\rm{dex}$ [statistical and systematic]; see \citealp{Soderblom09}), the metallicity of Praesepe is supersolar. Literature values for the absolute difference between the metallicity of the Pleiades and Praesepe vary from $\Delta \rm{[Fe/H]} =0.07-0.09\,\rm{dex}$ (see \citealp{An07} and \citealp*{Boesgaard13} respectively). We therefore adopt the mean value ($0.08\,\rm{dex}$) which translates to $Z_{\rm{Praesepe}}=1.2Z_{\rm{Pleiades}}$.

\begin{figure*}
\centering
\includegraphics[width=0.7\textwidth]{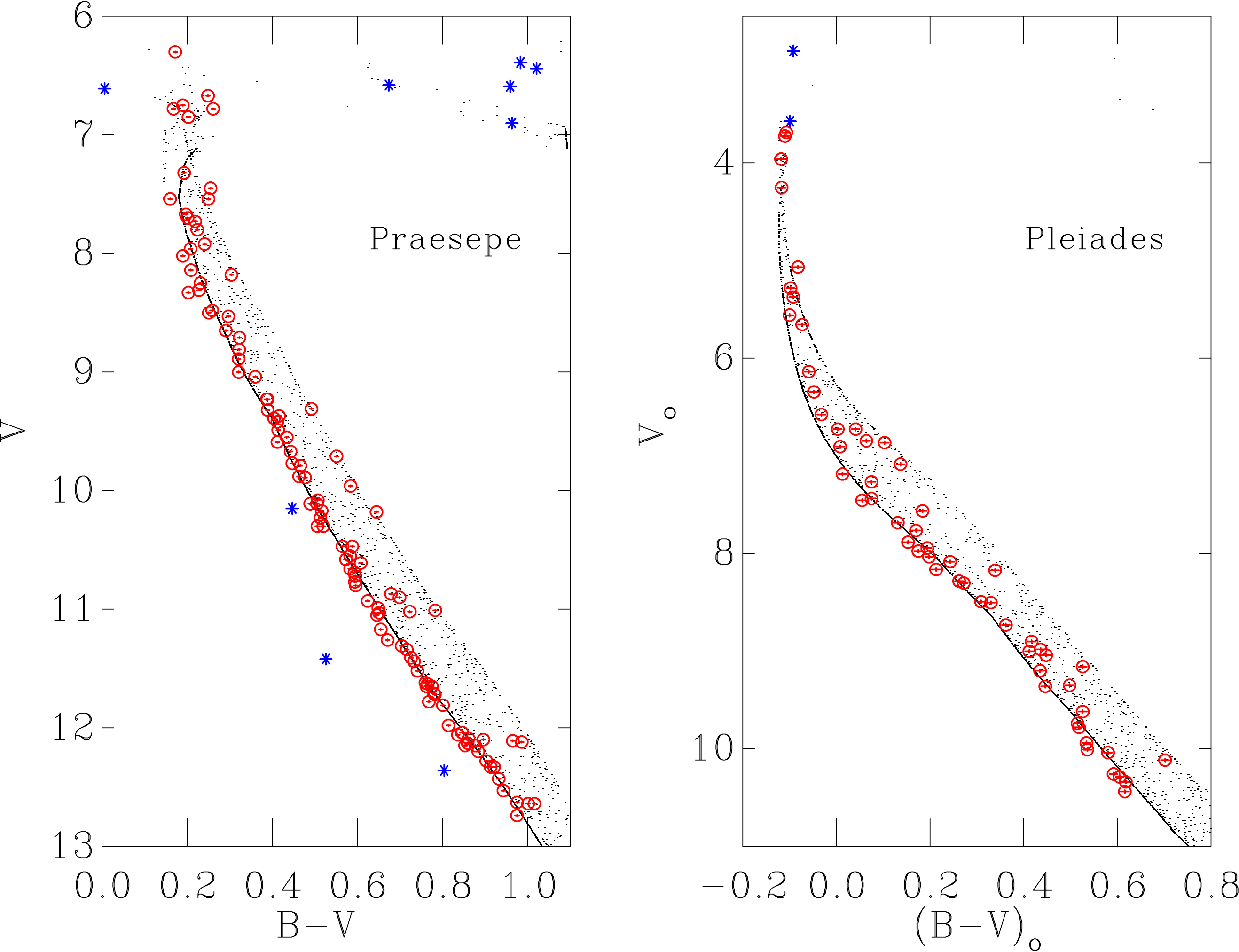}
\caption[]{Best-fitting CMDs of Praesepe and the Pleiades. \textbf{Left:} The best-fitting $V, B-V$ CMD for Praesepe
  with a derived age of $665\, \rm{Myr}$ and
  distance modulus $dm=6.32\,\rm{mag}$. Circles represent the data of
  \protect\cite{Johnson52}, with the associated uncertainties shown as
  the bars. Asterisks represent stars that were clipped before
  deriving the best-fit (see text). \textbf{Right:} The best-fitting de-reddened $V_{\circ}, (B-V)_{\circ}$ CMD for the Pleiades with an age of $135\,\rm{Myr}$ and distance modulus $dm=5.61\,\rm{mag}$. The data and uncertainties are from \protect\cite{Johnson53}.}
\label{fig:ms_cmd}
\end{figure*}

We performed a quadratic interpolation for both the interior and atmospheric models to produce grids of each with a metallicity equal to $1.2Z_{\odot}$. We note that at this metallicity the choice of interpolation scheme does not significantly affect the derived best-fit values for the age and distance i.e. adopting a linear interpolation we find that the the values of parameters are only affected at the $\sim 1\sigma$ level. To transform the theoretical models into CMD space, we reddened the atmospheric models by a nominal $E(B-V)=0.027\,\rm{mag}$ using the Galactic reddening law of \cite{Fitzpatrick99} with $R_{V}=3.1$ and then calculated the BCs as described in Section~\ref{atmospheric_models} using the \cite{Bessell12} system responses.

We used the  $\tau^{2}$ fitting statistic \citep{Naylor06,Naylor09} to fit the MS evolutionary models to the $UBV$ photometric data of Praesepe from \cite{Johnson52}. The reader is referred to Paper~II for a complete description of this process including the creation of the two-dimensional distributions and how we include the effects of binaries in the simulated population. The only difference is that as opposed to the power law distribution adopted in Paper~II, which results in a roughly even distribution of stars as a function of magnitude, we instead opt for the canonical broken power law IMF presented in \cite*{Dabringhausen08}. Even though we have adopted a different mass function for this study, we note that the choice of mass function does not have a significant impact on the derived best-fit parameters (see \citealp{Naylor09}) and hence the MS parameters we derive here are on a consistent scale with the MS parameters we have derived in Paper~II. If the assigned secondary mass is less than the lower mass limit of the interior models (which in the case of the \citealp{Schaller92} models is $0.8\,\msun$), we begin to lose the binary population before the single-star sequence in our two-dimensional distribution, resulting in a ``binary wedge" (see
\citealp{Jeffries07b}). To avoid this affecting our fitting of the
photometric data we combine the \cite{Schaller92} models with those of
\cite{Dotter08}, which extend to masses of $0.1\,\msun$. Between
$1\,\msun$ and $0.8\,\msun$ the two models show excellent agreement,
and so we use the \cite{Schaller92} models down to $1\,\msun$ and
append the \cite{Dotter08} models from $1\,\msun$ to $0.1\,\msun$.

The best-fitting $V, B-V$ CMD for Praesepe is shown in Fig.~\ref{fig:ms_cmd}. Prior to fitting we removed four stars defined as giants in \cite{Johnson52} in addition to one star which appears to have evolved off the MS as these would impact our derived values. Furthermore, we removed four additional stars due to a combination of their
positions in the $V, B-V$ CMD and their associated $\tau^{2}$ values which lie blueward of the MS locus. From the remaining stars we derive an age of
$665^{+14}_{-7}\,\rm{Myr}$ and a distance modulus $dm=6.32 \pm
0.02\,\rm{mag}$ with an
associated $\Pr(\tau^{2})=0.87$ (statistical uncertainties only).
As noted in Paper~II, the \cite{Schaller92} models do not include the pre-MS evolutionary phase, but only that from the ZAMS onwards. The age we derive for Praesepe is driven by the most massive stars i.e. those have evolved significantly away from the ZAMS. These most massive stars in Praesepe typically have spectral types of $\sim$ A5 (see e.g. \citealp{Bidelman56}, which, from the evolutionary models of \cite{Siess00} take $\sim 5\,\rm{Myr}$ to reach the ZAMS. Hence, although the \cite{Schaller92} models do not include the pre-MS evolutionary phase, the timescale involved is shorter than the uncertainties on the derived age and can therefore be ignored.

The values that we have derived for
the age and distance to Praesepe are consistent with, yet more precise than, previous
estimates e.g. $700\pm100\,\rm{Myr}$ \citep*{Salaris04}, $590^{+150}_{-120}\,\rm{Myr}$ \citep{Fossati08} and $dm=6.30 \pm 0.07$ \citep{VanLeeuwen09}. We note, however, that our derived age is significantly younger than that of \cite{Gaspar09}, who find an age of $757 \pm 36\,\rm{Myr}$ via isochrone fitting. 
\cite{Gaspar09} use SDSS photometry, however for bright stars the photometric measurements can be unreliable due to saturation effects. They therefore replace some of these measurements with transformed Johnson $BV$ data, however only when the difference between the transformed value and the original SDSS measurement is greater than $0.5\,\rm{mag}$. Thus, compared to the level of calibration and consistency in the Johnson $BV$ photometry we have used here, it is plausible that the age discrepancy could be due to residual unreliable photometry in the \cite{Gaspar09} sample.

We were concerned that different techniques had been used to derive the ages and distances for Praesepe and the Pleiades, which may be systematically different. As a check we therefore determined the Pleiades parameters using the same technique we had used for Praesepe. Starting with the photometric data of \cite{Johnson53}, we de-reddened the stars blueward of $B-V=0.0$ on a star-by-star basis using the revised Q-method (see \citealp{Mayne08}; Paper~II) as there was evidence of spatially variable extinction in the $U-B, B-V$ colour-colour diagram. Stars redward of $B-V=0.0$ were de-reddened using the same reddening vectors assuming a median reddening of $E(B-V)=0.03\,\rm{mag}$ (as calculated from the bluer stars). We fitted the entire dataset, deriving an age of 
$135^{+20}_{-11}\,\rm{Myr}$ and distance modulus $dm=5.61^{+0.03}_{-0.02}\,\rm{mag}$ with an associated $\Pr(\tau^{2})=0.23$ (see Fig.~\ref{fig:ms_cmd}; cf. $130\,\rm{Myr}$ and $dm=5.63$). Note that to achieve this fit it was necessary to remove the two brightest stars from our sample. The reason for this may simply be that we do not have enough stars in our simulated distribution to get enough density in that region of the CMD. Regardless, we can be reasonably confident that although we are using a different fiducial
locus in the SDSS photometric system, the adopted parameters for
both fiducial clusters are on a consistent age and distance scale.

\section{Creating semi-empirical pre-MS isochrones}
\label{creating_semi-empirical_pre-MS_isochrones}

The problems associated with using pre-MS model isochrones to derive
ages and masses from photometric data have been well documented over
the past two decades. It is strongly believed that such problems stem from underlying uncertainties in both the stellar interior and atmospheric models, especially at younger ages and lower $T_{\rm{eff}}$ (e.g. \citealp{Stauffer98a,Baraffe02,Hillenbrand04,Mayne07,DaRio10a}). In
Paper~I we demonstrated that there are severe problems with
using atmospheric models to transform pre-MS evolutionary models into
CMD space, where for $T_{\rm{eff}} \lesssim 4300\,\rm{K}$ the \emph{combined} models
overestimate the flux in the optical by up to a factor of two. Hence, if we
are to use pre-MS model isochrones to derive consistent ages and
masses for young stellar populations using CMDs, it is clear that we
must adopt empirical BCs for $T_{\rm{eff}}$ lower
than $\simeq 4300\,\rm{K}$.

\subsection{Deriving semi-empirical BC-$T_{\rm{eff}}$ relations}
\label{semi-empirical_bolometric_corrections}

\begin{table*}
\caption[]{The Pleiades single-star sequence in the combined Johnson-Cousins, 2MASS and INT-WFC bandpasses. Note these are apparent magnitudes and hence independent of the assumed distance and reddening. The full table is available as Supporting Information with the online version of the paper; a sample is shown here as a representation of its content. Note that the single-star sequence extends further in the $(RI)_{\rm{c}}$ and $(iZ)_{_{\rm{WFC}}}$ bandpasses (compared to the bluer and 2MASS bandpasses). Hence there becomes a point along the sequence where we are unable to tie both the $(RI)_{\rm{c}}$ and $(iZ)_{_{\rm{WFC}}}$ values to the $K_{\rm{s}}$-band values. Due to significant colour effects between the two photometric systems, at this point we choose to simply extend the $(iZ)_{_{\rm{WFC}}}$ sequence, but break the $(RI)_{\rm{c}}$ sequence and append the remaining values to the end of the table.}
\centering
\begin{tabular}{c c c c c c c c c c c}
\hline
$B$&$\gwfc$&$V$&$\rwfc$&$R_{\rm{c}}$&$\iwfc$&$I_{\rm{c}}$&$\Zwfc$&$J$&$H$&$K_{\rm{s}}$\\
\hline
11.384 & 10.983 & 10.700 & 10.514	& 10.322 & 10.398 &   9.957 & 10.372 &   9.482 &  9.177 &  9.076\\
11.501 & 11.091 & 10.800 & 10.609	& 10.414 & 10.483 & 10.042 & 10.450 &   9.559 &  9.243 &  9.138\\
11.619 & 11.198 & 10.900 & 10.702	& 10.504 & 10.567 & 10.126 & 10.528 &   9.633 &  9.306 &  9.198\\
11.737 & 11.302 & 11.000 & 10.793	& 10.595 & 10.649 & 10.210 & 10.604 &   9.705 &  9.368 &  9.256\\
11.854 & 11.407 & 11.100 & 10.882	& 10.684 & 10.729 & 10.294 & 10.678 &   9.775 &  9.428 &  9.313\\
11.973 & 11.508 & 11.200 & 10.969	& 10.774 & 10.807 & 10.376 & 10.750 &   9.841 &  9.484 &  9.367\\
12.092 & 11.607 & 11.300 & 11.053	& 10.862 & 10.883 & 10.458 & 10.820 &   9.904 &  9.538 &  9.419\\
12.210 & 11.708 & 11.400 & 11.139	& 10.951 & 10.959 & 10.540 & 10.890 &   9.966 &  9.591 &  9.470\\
12.329 & 11.806 & 11.500 & 11.220	& 11.038 & 11.032 & 10.620 & 10.957 & 10.024 &  9.640 &  9.518\\
12.448 & 11.907 & 11.600 & 11.304	& 11.125 & 11.106 & 10.699 & 11.025 & 10.082 &  9.689 &  9.565\\
\hline
\end{tabular}
\label{tab:pleiades_johnson-cousins}
\end{table*}

\begin{table}
\caption[]{The Praesepe single-star sequence in the SDSS $griz$ and 2MASS $K_{\rm{s}}$ bandpasses. Note these are apparent magnitudes and hence independent of the assumed distance and reddening. The full table is available as Supporting Information with the online version of the paper; a sample is shown here as a representation of its content.}
\centering
\begin{tabular}{c c c c c}
\hline
$g$&$r$&$i$&$z$&$K_{\rm{s}}$\\
\hline
15.330&14.009&13.467&13.156&11.143\\
15.385&14.056&13.501&13.182&11.166\\
15.440&14.102&13.535&13.208&11.190\\
15.496&14.148&13.570&13.234&11.213\\
15.551&14.194&13.604&13.260&11.236\\
15.606&14.241&13.638&13.286&11.259\\
15.661&14.287&13.673&13.312&11.282\\ 
15.717&14.333&13.707&13.338&11.305\\
15.772&14.379&13.742&13.364&11.328\\ 
15.827&14.426&13.776&13.390&11.351\\ 
\hline
\end{tabular}
\label{tab:praesepe_sdss}
\end{table}

To define the locus in each of the photometric systems, we fitted a spline (by eye) to the single-star sequence of the Pleiades for the Johnson-Cousins, 2MASS, INT-WFC and systems and Praesepe for the SDSS. To determine the position of the spline in CMD space we only used stars with uncertainties in both colour and magnitude of $\le 0.1\,\rm{mag}$. The splines are fitted as described in Paper~I and positioned taking into account the fact that the equal-mass binary locus lies $\simeq 0.75\,\rm{mag}$ above the single-star locus (with very few systems of higher multiplicity). The single-star sequence for the Pleiades in the combined Johnson-Cousins, 2MASS and INT-WFC bandpasses is given in Table~\ref{tab:pleiades_johnson-cousins}, whereas the Praesepe sequence in the SDSS bandpasses is given in Table~\ref{tab:praesepe_sdss}. The Pleiades single-star sequence in the IPHAS/UVEX system can be created by adopting the magnitudes in the INT-WFC bandpasses from Table~\ref{tab:pleiades_johnson-cousins} and converting these according to the transformations given in Table~\ref{tab:int-wfc_iphas-uvex}.

We have made a small modification to the single-star sequence in the $K_{\rm{s}}, g_{_{\rm{WFC}}}-K_{\rm{s}}$ CMD (in comparison to that given in Paper~I) in the colour range where we suffered from a paucity of stars ($g_{_{\rm{WFC}}}-K_{\rm{s}} \simeq 4.5-5.5$) in the original definition. As a result of additional Pleiades (Rees et al. in preparation) and Praesepe observations (see Section~\ref{comparing_pleidaes_praesepe_cmds}) this region is now better sampled. This modification does not affect the resultant semi-empirical model isochrones in the optical INT-WFC colours, and therefore does not change the conclusions of Paper~II. Furthermore, it also ensures that any isochrones involving the $K_{\rm{s}}$-band are now more representative than previous formulations.

We note that although defining a locus by eye is somewhat of a subjective process, the cluster sequence in each photometric system is sufficiently populated and well-defined that the uncertainties introduced into the position of the single-star locus are minimal. The overriding source of uncertainty comes from the membership catalogues we have adopted to select cluster members. However, as these are based on a combination of photometric and kinematic diagnostics, coupled with the fact that we have selected members with high probabilities, we are confident that the single-star sequences given in Tables~\ref{tab:pleiades_johnson-cousins} and \ref{tab:praesepe_sdss} are accurate to within a few hundredths of a magnitude.

Comparing our Pleiades single-star sequence in the $BVI_{\rm{c}}K_{\rm{s}}$ bandpasses to that of \cite{Stauffer07}, in addition to an updated formalism in \citealp{Kamai14}, (both of whom also defined the single-star sequence by eye), we notice that our definition is almost identical to both across all bandpasses, expect for small regions where we find differences of the order of $<0.05\,\rm{mag}$ in colour at a given magnitude. We note that such differences are \emph{not} systematic over the entire colour range of the Pleiades dataset, but are instead localised to regions with a relative paucity of stars. We can therefore estimate what effect such localised differences in the single-star sequence has on our age scale by translating the difference in colour at a given magnitude into a fractional difference in age. For a difference (e.g. in the $V-I_{\rm{c}}$ colour) of $0.05\,\rm{mag}$ we find that this equates to a difference in age of $\lesssim10$ per cent. Hence, this represents the systematic uncertainty in our age scale (within specific narrow colour ranges) as a result of adopting our own definition of the Pleiades single-star sequence instead of others available in the literature.

\begin{figure}
\centering
\includegraphics[width=\columnwidth]{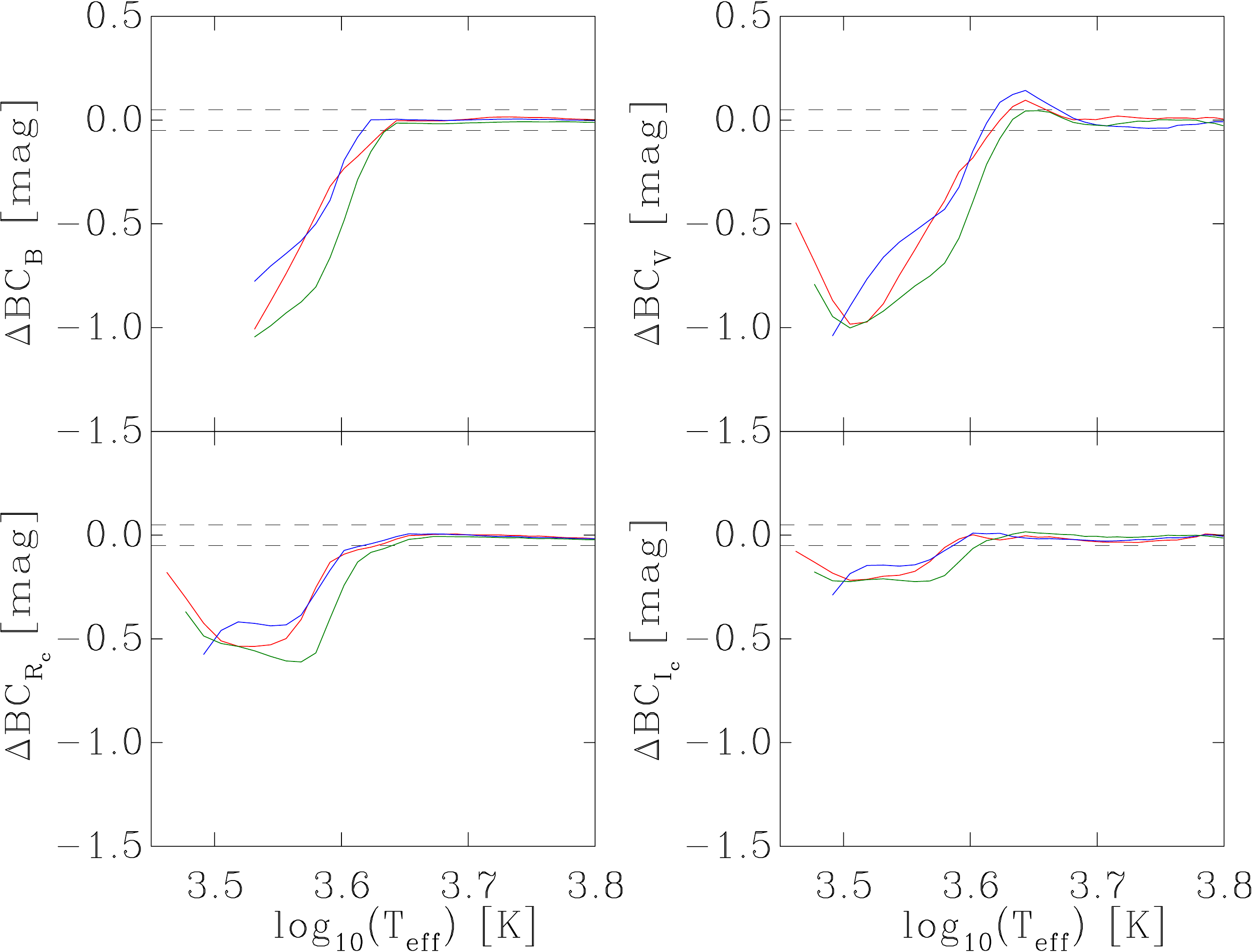}
\caption[]{Model-dependent corrections ($\Delta$BC; defined as $\rm{BC_{\rm{empirical}}}-\rm{BC_{\rm{theory}}}$) calculated as a function of $T_{\rm{eff}}$ for the optical $BV(RI)_{\rm{c}}$ Johnson-Cousins bandpasses. Corrections are shown for the following models: BCAH98 $\alpha=1.9$ (\emph{red}), DCJ08 (\emph{blue}) and updated Pisa (\emph{green}). The dashed lines represent the $\pm 0.05\,\rm{mag}$ level with respect to zero.}
\label{fig:delta_bc_johnson}
\end{figure}

In Papers~I and II we described the process of deriving empirical BCs at each point along the fiducial locus and then expressing this as a correction ($\Delta$BC) to the theoretical BC derived for the appropriate $T_{\rm{eff}}$ and log$\,g$ for that point in the sequence. In Paper~I we used binary systems with well-constrained dynamical masses to demonstrate that the models fit the $K_{\rm{s}}$-band flux well. We can then use this to set the $T_{\rm{eff}}$ at each point along our fiducial locus (this also ensures that we have a well-defined mass scale tied to that of the adopted sequence). Looking, for instance, at the $K_{\rm{s}}, V-K_{\rm{s}}$ CMD, the necessary $\Delta$BC$_{V}$ can be determined by calculating the difference between the fiducial locus and the model isochrone at a given $T_{\rm{eff}}$. Repeating this process over the colour range of the fiducial locus or $T_{\rm{eff}}$ range of the model isochrone (whichever is more restrictive) results in a set of $T_{\rm{eff}}$-dependent $\Delta$BC$_{V}$s. These corrections can then be added to the theoretical BC grid (which is a function of $T_{\rm{eff}}$ and log$\,g$) at the appropriate $T_{\rm{eff}}$ irrespective of its log$\,g$., thereby defining a set of log$\,g$-dependent semi-empirical BC$_{V}$s. This process can be extended to calculate model-dependent $\Delta$BCs in other bandpasses. \emph{As demonstrated in Paper~I, the models match the observed shape of the Pleiades locus at $T_{\rm{eff}} \gtrsim 4300\,\rm{K}$, and hence we only apply $\Delta$BCs to the theoretical BC-$T_{\rm{eff}}$ relation below this $T_{\rm{eff}}$}. As an illustration, in Fig.~\ref{fig:delta_bc_johnson} we show the derived model-dependent $\Delta$BCs in the Johnson-Cousins photometric system. A full discussion on the quantification of the observed discrepancy and the effect this has on the derived ages for young stars in this $T_{\rm{eff}}$ regime is given in Paper~I.

\begin{figure*}
\centering
\includegraphics[width=\textwidth]{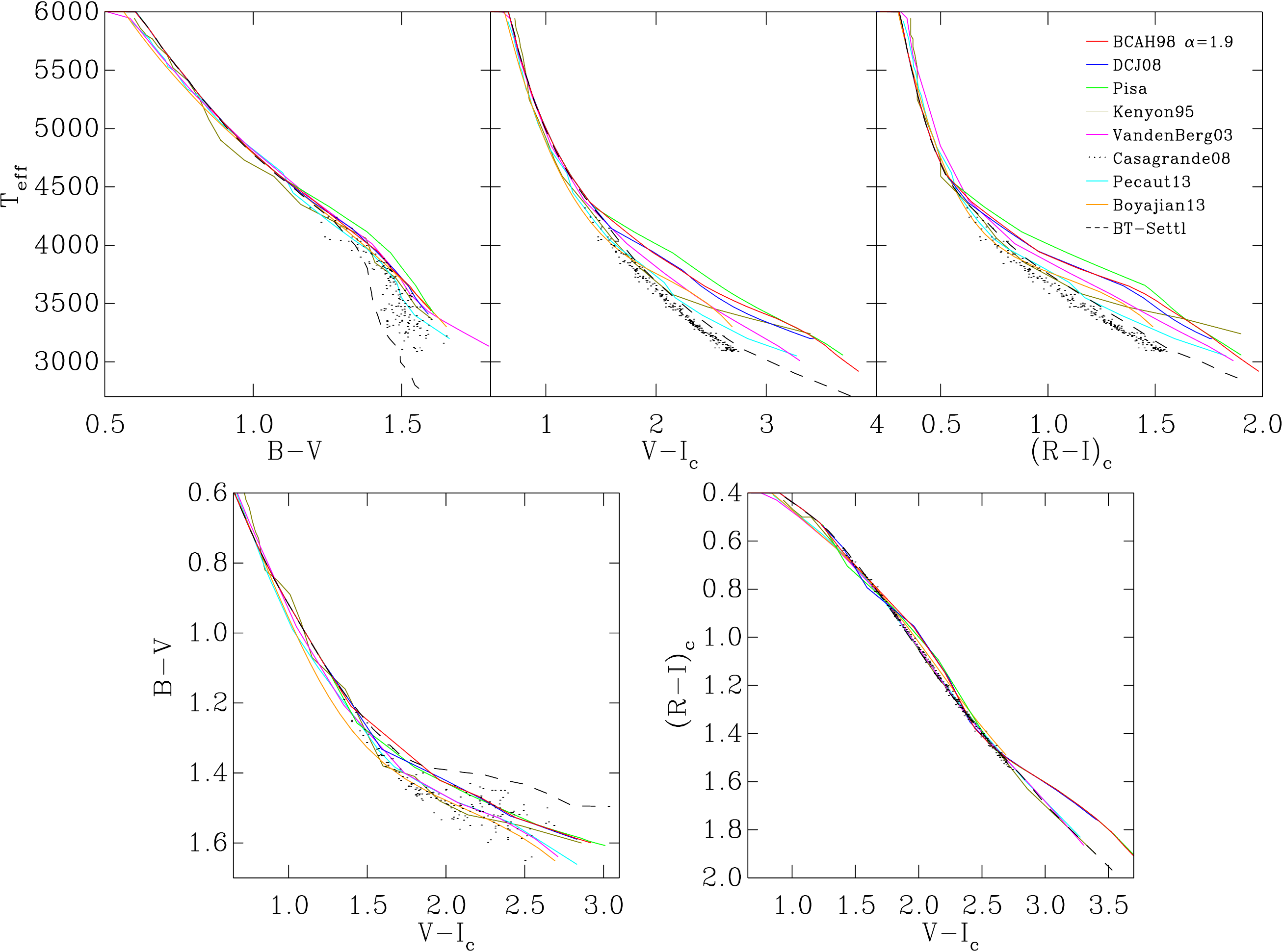}
\caption[]{Colour-$T_{\rm{eff}}$ and colour-colour relations in the Johnson-Cousins
  photometric system created using semi-empirical pre-MS model
  isochrones at an age of $1\,\rm{Gyr}$.
  In each panel the relations derived using the BCAH98
  $\alpha=1.9$ (\emph{red}), DCJ08 (\emph{blue}) and updated Pisa models (\emph{green}) are
  shown in addition to the empirical relations of \cite{Kenyon95} (dwarf stars; \emph{olive}), \cite{Vandenberg03} (dwarf stars; \emph{magenta}), \cite{Casagrande08} (dwarf stars with mixed metallicity; \emph{black dots}), \cite{Pecaut13} (dwarf stars; \emph{cyan}), and \cite{Boyajian13} (dwarf stars; \emph{orange}). Note that the \cite{Casagrande08} sample represents another empirical formalism, however due to metallicity effects (especially in the $B-V$ colour) a polynomial was not fitted through the sequence, and thus we retain them as individual points. For reference we also plot the theoretical relation derived using the BT-Settl models at a log$\,g=4.5$ (\emph{black dashed}).}
\label{fig:colour_teff}
\end{figure*}

This process was repeated for each of the three sets of evolutionary models in the four photometric systems (we create the corresponding IPHAS/UVEX semi-empirical BC-$T_{\rm{eff}}$ relations by applying the transformations given in Table~\ref{tab:int-wfc_iphas-uvex} to the derived relations in the INT-WFC system). Both the BCAH98 $\alpha=1.9$ and the updated Pisa models are only available in solar metallicities, however Praesepe has a slightly supersolar composition (see Section~\ref{praesepe_parameters}). Thus, for these models we simply overlay supersolar atmospheres (1.2$Z_{\odot}$) on the solar composition interior models and calculate the corrections to the BCs as described above. We shall consider the effects of this in Section~\ref{effects_metallicity_semi-empirical_BC-Teff_relations}. The DCJ08 models, however, are available in supersolar metallicities, so for these we perform a quadratic interpolation (as in Section~\ref{praesepe_parameters}) to create a grid of 1.2$Z_{\odot}$ interior models and then combine these with the 1.2$Z_{\odot}$ atmospheric models.
Note that as we have used the $K_{\rm{s}}$-band
magnitude of the pre-MS model isochrones as a proxy for
$T_{\rm{eff}}$, each set of pre-MS interior models requires a slightly
different set of semi-empirical BCs to fit the fiducial
locus in CMD space. This is due to variations in the underlying physical inputs and
assumptions in the models, however, we find that the resultant semi-empirical BC-$T_{\rm{eff}}$ relations agree to within $0.15\,\rm{mag}$ at a given $T_{\rm{eff}}$ across all colours (see e.g. Fig.~\ref{fig:colour_teff}).

Due to exposure time limits for individual fields-of-view in the SDSS, the Praesepe locus only enables us to recalibrate the colour-$T_{\rm{eff}}$ relation to a limit of $g-i \simeq 3.5$. In comparison, our Pleiades locus in the INT-WFC photometric system extends to $\giwfc \simeq 4.0$. The $\Delta$BCs for each system are significantly different at $\simeq 3900\,\rm{K}$ (e.g. $\Delta\mathrm{BC}_{g}-\Delta\mathrm{BC}_{g_{_{\rm{WFC}}}} \sim 0.1\,\rm{mag}$), however they converge towards $3400\,\rm{K}$ (e.g. $\Delta\mathrm{BC}_{g}-\Delta\mathrm{BC}_{g_{_{\rm{WFC}}}}=0.04\,\rm{mag}$). Thus, to extend our semi-empirical model isochrones to lower $T_{\rm{eff}}$ in the SDSS system we adopt the $\Delta$BCs we have derived for the INT-WFC system and apply these to the equivalent SDSS $griz$ bandpasses. Although this is not an ideal solution, it does allow us to extend the semi-empirical SDSS model isochrones to $T_{\rm{eff}} \simeq 3100\,\rm{K}$, which is a significant advantage and certainly constitutes a marked improvement over adopting purely theoretical model isochrones at such $T_{\rm{eff}}$.

\subsection{Comparison with empirical colour-$T_{\rm{eff}}$ relations}
\label{comparison_with_literature_relations}

Having derived a set of log$\,g$-dependent semi-empirical BC-$T_{\rm{eff}}$ relations in Section~\ref{semi-empirical_bolometric_corrections} we can now compare these to empirical colour-$T_{\rm{eff}}$ relations available in the literature. The Johnson-Cousins system is, at least historically, the most commonly used photometric system and as such there are several independent relations with which to compare against. Given that the majority of the empirical relations in the literature are based on observations of MS stars, we therefore adopt semi-empirical isochrones with ages of $1\,\rm{Gyr}$ for this comparison. Even though we derived the $\Delta$BCs in the Johnson-Cousins system using the observed colours of Pleiades members, and hence pre-MS objects at the lowest masses, we have applied the derived $\Delta$BC at a specific $T_{\rm{eff}}$ to all values of log$\,g$ at that $T_{\rm{eff}}$ (see Assumption 2 in Section~\ref{intro:methodology}). Thus we can create semi-empirical isochrones across a wide range of ages, and subsequently log$\,g$, and not just a narrow range close to that of the fiducial cluster.

Fig.~\ref{fig:colour_teff} shows how the semi-empirical BC-$T_{\rm{eff}}$ relations, as defined by semi-empirical model isochrones with ages of $1\,\rm{Gyr}$, compare with several well-known empirical relations, namely those of \cite{Kenyon95}\footnote[4]{These relations in fact constitute a compilation of adopted relations from the then current literature and were taken from \cite{Johnson66}, \cite{Schmidt82}, \cite{Bessell88}, and \cite{Bessell90a}.}, \cite{Vandenberg03}, \cite*{Casagrande08}, \cite{Pecaut13}, and \cite{Boyajian13}. As a reference, in Fig.~\ref{fig:colour_teff} we also plot the theoretical colour-$T_{\rm{eff}}$ and colour-colour relations predicted by the \textsc{phoenix} BT-Settl models for log$\,g=4.5$. We note that the \cite{Casagrande08} sample represents another empirical formalism, and should not be considered as data against which the various colour-$T_{\rm{eff}}$ and colour-colour relations shown in Fig.~\ref{fig:colour_teff} should be compared against. Due to metallicity effects (especially in the $B-V$ colour, see below) a polynomial was not fitted through the sequence, and thus we retain them as individual points in Fig.~\ref{fig:colour_teff}.

From the upper panels of Fig.~\ref{fig:colour_teff} it is clear that, even amongst the empirical relations, in the low temperature regime there is considerable scatter in the colour one would adopt for a given $T_{\rm{eff}}$ (as large as $1\,\rm{mag}$ in $V-I_{\rm{c}}$ at $T_{\rm{eff}} \simeq 3200\,\rm{K}$). What is noticeable, however, is that the relations of \cite{Vandenberg03} and \cite{Pecaut13} appear to most closely match the observed shape of the \cite{Casagrande08} sample, whereas the \cite{Boyajian13} relation appears to lie somewhere in-between these two formalisms and the \cite{Kenyon95} relation diverges [especially in the $V-I_{\rm{c}}$ and $(R-I)_{\rm{c}}$ planes] from the other empirical relations at $T_{\rm{eff}} \lesssim 3500\,\rm{K}$. Furthermore, it is evident that the spread in the \cite{Casagrande08} sample is significantly larger in the $B-V$ plane compared to the $V-I_{\rm{c}}$ and $(R-I)_{\rm{c}}$ planes which is suggestive of a strong metallicity effect in the $B-V$ colour index that is not present at longer wavelengths (see also \citealp{Ramirez05,Casagrande10}). Also evident from Fig.~\ref{fig:colour_teff} is that the theoretical $B-V$ colour-$T_{\rm{eff}}$ relation from the BT-Settl models is too blue compared to the other empirical relations.

The colour-colour relations plotted in the lower panels of Fig.~\ref{fig:colour_teff} imply that whilst the colour scales of these empirical relations are likely reliable, constraining the $T_{\rm{eff}}$ scale, especially at low temperatures, is much more problematic. This fact is reflected in the various methods which have been adopted to derive the $T_{\rm{eff}}$ scales for cool stars in these relations, including visually fitting blackbody curves to observed data (\citealp{Kenyon95}, see also \citealp{Johnson66}), a modified infrared flux method \citep{Casagrande08}, multi-band spectral energy distribution fitting \citep{Pecaut13} and direct measurement of angular diameters \citep{Boyajian13}. It is interesting to note that despite the significant differences between the various empirical $V-I_{\rm{c}}$ and $(R-I)_{\rm{c}}$ colour-$T_{\rm{eff}}$ relations, the $V-I_{\rm{c}}$, $(R-I)_{\rm{c}}$ colour-colour relation shows remarkable agreement between the various formalisms.

Although our colour-$T_{\rm{eff}}$ relations are much redder than the other relations shown in Fig.~\ref{fig:colour_teff}, the relations we have derived are those required to transform a given pre-MS interior model so as to follow the observed fiducial locus in various CMDs i.e. they represent the necessary corrections to make the models work better. As we have combined both the interior and atmospheric models to calculate theoretical colours and magnitudes, the derived discrepancy between the models and the data cannot simply be attributed to one or the other. Instead it is a combination of the uncertainties at low $T_{\rm{eff}}$ associated with both sets of models, and likely includes contributions from the atmospheric models (e.g. incomplete opacity sources) and interior models (e.g. treatment of convection) in addition to physical processes which affect the data, but are not included in the models (e.g. magnetic activity).

\begin{figure*}
\centering
\includegraphics[width=\textwidth]{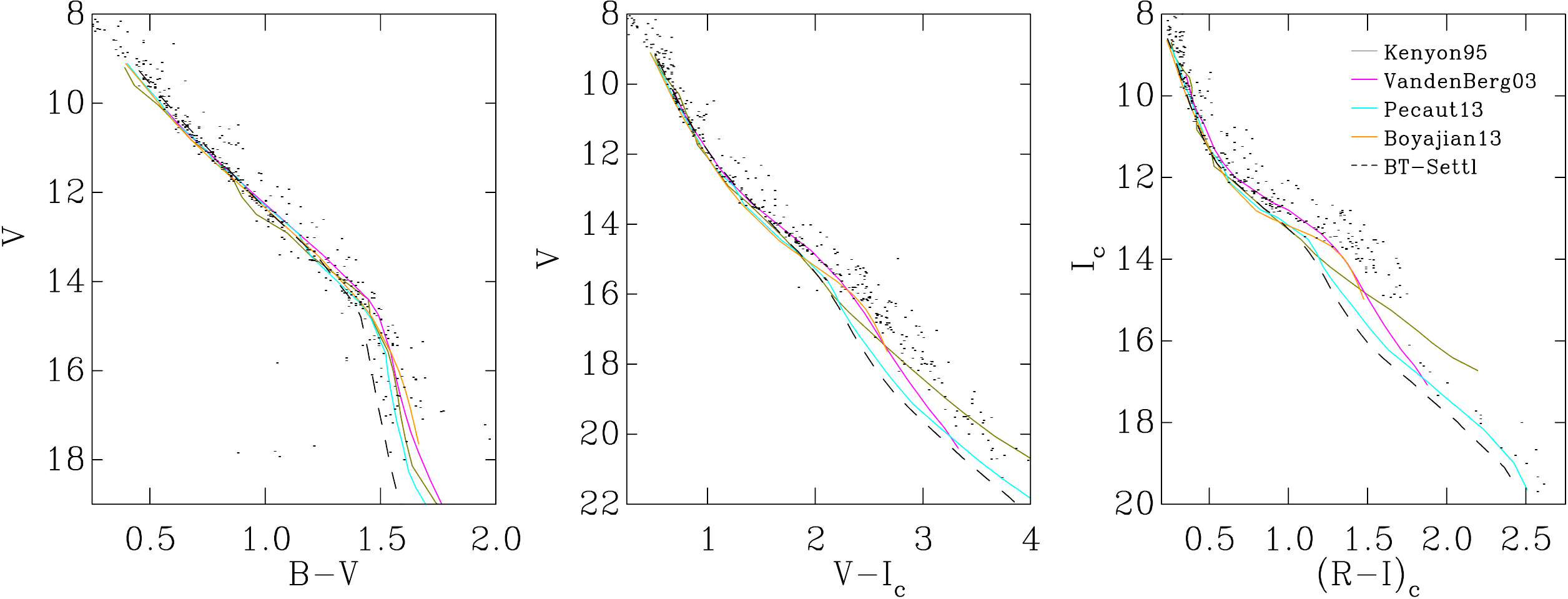}
\caption[]{Optical CMDs of Pleiades members in the Johnson-Cousins photometric system. \textbf{Left:} $V, B-V$ CMD. \textbf{Middle:} $V, V-I_{\rm{c}}$ CMD. \textbf{Right:} $I_{\rm{c}}, (R-I)_{\rm{c}}$ CMD. Overlaid are $130\,\rm{Myr}$ BCAH98 $\alpha=1.9$ pre-MS model isochrones transformed into CMD space using the empirical BCs and colour-$T_{\rm{eff}}$ relations of \cite{Kenyon95} (\emph{olive}), \cite{Vandenberg03} (\emph{magenta}), \cite{Pecaut13} (\emph{cyan}), and \cite{Boyajian13} (\emph{orange}). Note that \cite{Boyajian13} do not provide empirical BCs with their colour-$T_{\rm{eff}}$ relations and therefore we adopt the dwarf BC$_{V}$-$T_{\rm{eff}}$ relation of \cite{Pecaut13}. In each panel, we have adopted a distance modulus $dm=5.63\,\rm{mag}$ and a reddening of $E(B-V)=0.04\,\rm{mag}$ [equivalent to $E(V-I_{\rm{c}})=0.05\,\rm{mag}$ and $E(R_{\rm{c}}-I_{\rm{c}})=0.03\,\rm{mag}$]. For reference we also plot a $130\,\rm{Myr}$ BCAH98 $\alpha=1.9$ model isochrone transformed using the BT-Settl atmospheric models (\emph{black dashed}).}
\label{fig:pleiades_cmd}
\end{figure*}

In Fig.~\ref{fig:pleiades_cmd} we further demonstrate the observed scatter between the different empirical colour-$T_{\rm{eff}}$ relations by using these to transform a $130\,\rm{Myr}$ BCAH98 $\alpha=1.9$ model isochrone into various optical CMDs and compare these against our Pleiades catalogue (described in Section~\ref{data:johnson_cousins}). For this comparison we adopted the distance and reddening values stated in Section~\ref{pleiades_parameters}. Note that \cite{Boyajian13} do not provide empirical BCs with their colour-$T_{\rm{eff}}$ relations and therefore we adopt the dwarf BC$_{V}$-$T_{\rm{eff}}$ relation of \cite{Pecaut13}.

It is clear that of the empirical relations tested, those of \cite{Vandenberg03} provide the closest match to the observed shape of the Pleiades locus, followed by those of \cite{Boyajian13}, whilst those of \cite{Kenyon95} and \cite{Pecaut13} show significant deviations. It is interesting to note that all the relations appear to model the Pleiades locus similarly well in the $V, B-V$ CMD, whereas in all the CMDs the theoretical BT-Settl relations predict colours that are too blue (see also \citealp{Baraffe98} who demonstrated this effect for previous generations of {\sc{phoenix}} atmospheric models).

\section{Factors affecting the use of Praesepe in place of the Pleiades}
\label{praesepe_suitability}

In Section~\ref{data:sdss} we highlighted two possible factors which need to be considered when adopting Praesepe (as opposed to the Pleiades) as a fiducial cluster. First, there is the question of enhanced magnetic activity at younger ages i.e. when we compare the loci of the Pleiades and Praesepe in different CMDs do we see the same effects as noted by \cite{Stauffer03} (see also \citealp{Kamai14})? Secondly, what systematic residuals are introduced by using a stellar locus with a supersolar metallicity to recalibrate solar composition pre-MS model isochrones?

\subsection{Effects of enhanced magnetic activity}
\label{effects_of_enhanced_magnetic_activity}

Enhanced levels of magnetic activity, especially amongst younger stars, can inhibit the convective flows to the stellar surface and result in star spots covering a large fraction of the photosphere \citep{Strassmeier09}. Depending on the areal coverage of such spots, the observed colours can be somewhat different from older stars of the same $T_{\rm{eff}}$. An additional consequence of the inhibited convective flows is that the radii of stars with intense magnetic fields can be significantly inflated in comparison to stars of a similar mass but with a much weaker field (e.g. \citealp*{Chabrier07}; \citealp{Yee10}). It is beyond the scope of this paper to discuss these processes in too fine a detail as we are primarily concerned with whether they could introduce systematic differences into the positions of the Praesepe locus (when compared to that of the Pleiades) in the CMD, and therefore introduce possible biases into the calculated $\Delta$BCs. 

\cite{Stauffer03} identified a colour anomaly for K dwarfs in the Pleiades (see also \citealp{Jones72,Mermilliod92,Kamai14}) which is present in both the $V, B-V$ and $V, V-K$ CMDs, but vanishes in the $V, V-I$ CMD. They attributed these differences to a combination of rapid rotation and the presence of star spots on the stellar surface, both of which are observed to become less pronounced with age. Thus, given that we have adopted Praesepe as our fiducial cluster in the SDSS photometric system, we must therefore ask whether we still observe the effects noted by \cite{Stauffer03} in the equivalent $g, g-i$ and $g, g-K_{\rm{s}}$ CMDs.

\subsubsection{CMD comparison of the Pleiades and Praesepe}
\label{comparing_pleidaes_praesepe_cmds}

As discussed in Section~\ref{data:sdss}, Pleiades data is not available in the SDSS photometric system. Therefore, if we are to compare the sequences of the Pleiades and Praesepe in CMD space to see whether the effect noted by \cite{Stauffer03} is present in either the $g, g-i$ or $g, g-K_{\rm{s}}$ CMD, we must do so in the natural photometric system of the INT-WFC. Whilst there are differences between the SDSS and INT-WFC response functions (see Paper~I), they are similar enough that the conclusions based on a comparison in the INT-WFC colours will likely be valid for the SDSS colours.

\begin{table}
\caption[]{The central coordinates for each field-of-view and exposure times in the INT-WFC bandpasses for the observations of Praesepe.}
\centering
\begin{tabular}{c c c c}
\hline
Field&RA Dec.&Bandpass&Exposure time (s)\\
&(J2000.0)&&$\times 1$ unless stated\\
\hline
Praesepe&$08^{\rm{h}}40^{\rm{m}}46.7^{\rm{s}}$&$g_{_{\rm{WFC}}}$&1, 10($\times 3$), 100\\
Field~A&$+19^{\circ}32^{'}56.0^{''}$&$r_{_{\rm{WFC}}}$&1, 10($\times 3$)\\
&&$i_{_{\rm{WFC}}}$&1, 10($\times 3$), 100\\
Praesepe&$08^{\rm{h}}40^{\rm{m}}22.7^{\rm{s}}$&$g_{_{\rm{WFC}}}$&1, 10($\times 3$)\\
Field~B&$+19^{\circ}59^{'}59.6^{''}$&$r_{_{\rm{WFC}}}$&1, 10($\times 3$)\\
&&$i_{_{\rm{WFC}}}$&1, 10($\times 3$)\\
\hline
\end{tabular}
\label{table:praesepe}
\end{table}

Observations of two fields within Praesepe (see Table~\ref{table:praesepe}) were taken with the 2.5-m INT on La Palma on the nights of 2012 December 5 and 2013 November 11. We used the same instrumentation and filter set on both occasions, and we refer the reader to Paper~I for details on these as well as our observational techniques, photometric calibration, data reduction and astrometric calibration. Only the 2013 data were taken in photometric conditions and thus  standard star fields in Stripe~82 of the SDSS were observed throughout the night.
 To create our final optical catalogue, the 2012 and 2013 data were reduced separately and a normalisation process (see \citealp{Naylor02}; Paper~I) performed on the two separate catalogues. The zero-point shift required for each field (effectively allowing for small variations between fields observed on different nights) is a reliable indicator of the internal consistency of the photometry, as well as the accuracy with which the profile corrections was performed, and suggests an accuracy of better than 2 per cent in all bandpasses. Our full Praesepe photometric catalogue is given in Table~\ref{tab:praesepe_full}.

\begin{figure*}
\centering
\includegraphics[width=\textwidth]{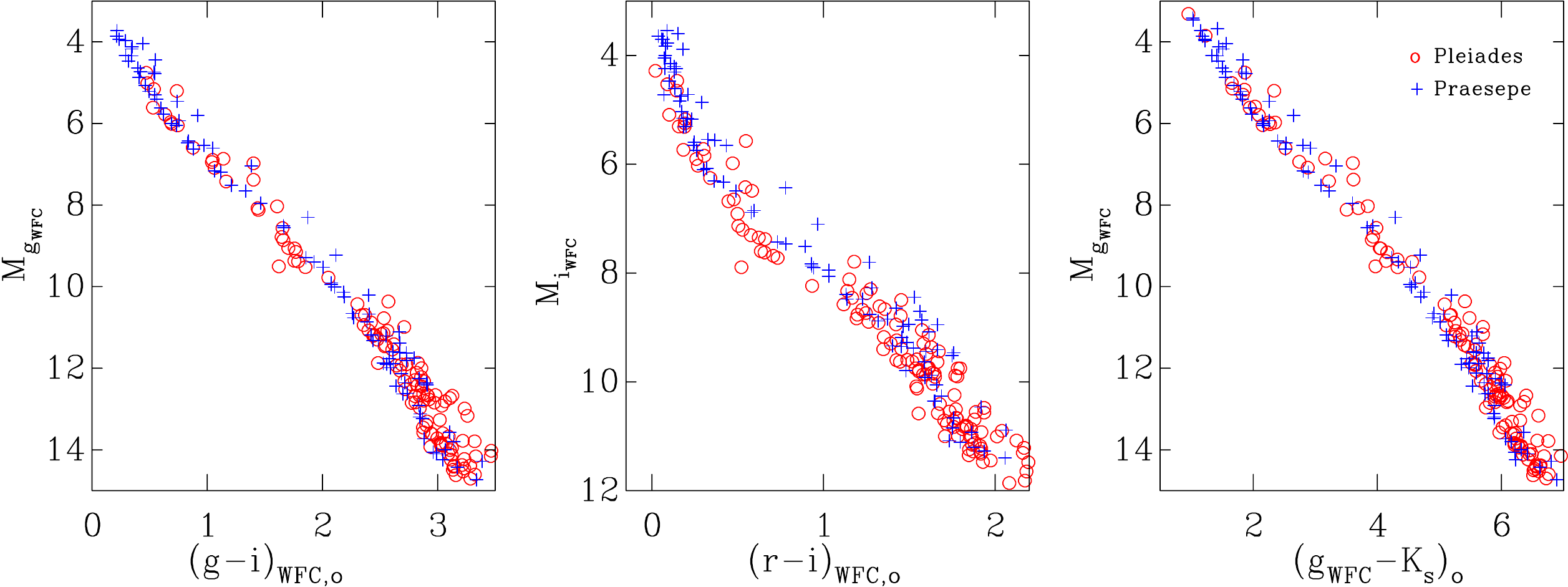}
\caption[]{Comparison of the Pleiades (\emph{red circles}) and Praesepe (\emph{blue crosses}) members in CMDs. \textbf{Left:} $M_{g_{_{\rm{WFC}}}}, (g-i)_{_{\rm{WFC,\circ}}}$ CMD. \textbf{Middle:} $M_{i_{_{\rm{WFC}}}}, (r-i)_{_{\rm{WFC,\circ}}}$ CMD. \textbf{Right:} $M_{g_{_{\rm{WFC}}}}, (g_{_{\rm{WFC}}}-K_{\rm{s}})_{\circ}$. There is no indication of systematic differences between the two loci at low masses in any of the CMDS.}
\label{fig:pleiades_praesepe}
\end{figure*}

\begin{table*}
\caption{A sample of the full Praesepe photometric catalogue with colours and magnitudes in the natural INT-WFC photometric system. Due to space restrictions, we only show
  the $\gwfc$ and $\giwfc$ colours and
  magnitudes as a representation of its content. The full photometric
  catalogue (available as Supporting Information with the online version of the paper) also includes photometry in the
  $\riwfc$ colour and the $\rwfc$ and $\iwfc$ magnitudes. Columns list unique
  identifiers for each star in the catalogue: field and CCD number (integer and decimal), ID, RA and
  Dec. (J2000.0), CCD pixel coordinates of the star, and for each of
  $g_{_{\rm{WFC}}}$ and $(g-i)_{_{\rm{WFC}}}$ there is a magnitude,
  uncertainty on the magnitude and a flag (OO represents a ``clean
  detection''; see \protect\citealp{Burningham03} for a full description of
  the flags.)}
\begin{tabular}{c c c c c c c c c c c c}
\hline
Field&ID&RA (J2000.0)&Dec. (J2000.0)&x&y&$\gwfc$&$\sigma_{
g_{_{\rm{WFC}}}}$&Flag&$\giwfc$&$\sigma_{
(g-i)_{_{\rm{WFC}}}}$&Flag\\
\hline
1.04&12&08 40 56.294&+19 34
49.25&602.610&1641.420&8.387&0.010&OS&-0.191&0.014&SS\\
1.02&4&08 39 56.450&+19 33
10.98&2082.593&813.918&8.415&0.010&OS&-0.094&0.014&SS\\
\hline
\end{tabular}
\label{tab:praesepe_full}
\end{table*}

We adopt the same \cite{Kraus07} members as described in Section~\ref{data:sdss} to isolate Praesepe members and we present these in the natural INT-WFC photometric system in Table~\ref{tab:praesepe_members}. For the Pleiades and Praesepe we use the distance and reddening values stated in Sections~\ref{pleiades_parameters} and \ref{praesepe_parameters} respectively. We derive the extinction in each bandpass for both clusters using the extinction grids which we discuss in detail in Section~\ref{reddening_extinction} and which are created using the atmospheric models, the INT-WFC system responses and the Galactic reddening law of \cite{Fitzpatrick99}. Note that over the colour range of the available photometric data, in addition to the low reddening affecting both clusters, the derived extinction values are insensitive to the adopted $T_{\rm{eff}}$ and log$\,g$, and are affected at only the few millimag level.

\begin{table*}
\caption{A sample of the catalogue of colours and magnitudes in the natural INT-WFC photometric system for Praesepe members. The columns and content are in the same format as Table~\ref{tab:praesepe_full}. The full photometric catalogue is available as Supporting Information with the online version of the paper.}
\begin{tabular}{c c c c c c c c c c c c}
\hline
Field&ID&RA (J2000.0)&Dec. (J2000.0)&x&y&$\gwfc$&$\sigma_{
g_{_{\rm{WFC}}}}$&Flag&$\giwfc$&$\sigma_{
(g-i)_{_{\rm{WFC}}}}$&Flag\\
\hline
2.02&12&08 39 03.549&+19 59
59.20&873.852&1047.167&9.174&0.010&OS&0.396&0.014&SS\\
2.03&22&08 40 52.437&+20 15
59.48&293.658&758.948&9.214&0.010&OS&0.415&0.014&SS\\
\hline
\end{tabular}
\label{tab:praesepe_members}
\end{table*}

Fig.~\ref{fig:pleiades_praesepe} compares the loci of the Pleiades and Praesepe in the $M_{g_{_{\rm{WFC}}}}, (g-i)_{_{\rm{WFC,\circ}}}$, $M_{i_{_{\rm{WFC}}}}, (r-i)_{_{\rm{WFC,\circ}}}$ and $M_{g_{_{\rm{WFC}}}}, (g_{_{\rm{WFC}}}-K_{\rm{s}})_{\circ}$ CMDs. The Pleiades members are those we identified in Paper~I. As noted in \cite{Stauffer03} (see also \citealp{Kamai14}), the low-mass Pleiades locus in the $V, B-V$ CMD appeared systematically bluer than that of Praesepe and redder in the $V, V-K_{\rm{s}}$ CMD. From Fig.~\ref{fig:pleiades_praesepe} we find no evidence of a significant difference between the two loci in any of the CMDs. This is rather surprising, especially in the $M_{g_{_{\rm{WFC}}}}, (g-i)_{_{\rm{WFC,\circ}}}$ CMD, given that the $\gwfc$-band has a significant wavelength overlap with the Johnson $B$-band. We note, however, that in the $M_{i_{_{\rm{WFC}}}}, (r-i)_{_{\rm{WFC,\circ}}}$ and $M_{g_{_{\rm{WFC}}}}, (g_{_{\rm{WFC}}}-K_{\rm{s}})_{\circ}$ CMDs there may be tentative evidence of small $< 0.1\,\rm{mag}$ differences in colour at a given magnitude. These differences are typically smaller than those discussed in the studies of \cite{Stauffer03} and \cite{Kamai14}, and are associated with regions in the loci of low stellar density. Additional photometric observations of both clusters would be required before we can ascertain whether these differences are indeed genuine variations between the two loci or simply selection effects due to our current photometric coverage of each cluster.

As highlighted in \cite{Stauffer03}, the anomalous $B-V$ colours may be due to an enhanced blue continuum, especially at wavelengths less than $\simeq 4200\,\rm{\AA}$. Hence, the reason we may not be observing this effect could be due to the fact that, compared to the $\gwfc$-band, the Johnson $B$-band is bluer (blue wavelength cut-off $\simeq 3600\,\rm{\AA}$, cf. $\simeq 4000\,\rm{\AA}$), more skewed to the blue (maximum throughput at $\simeq 4000\,\rm{\AA}$, cf. $\simeq 5000\,\rm{\AA}$) and somewhat narrower ($\simeq 900\,\rm{\AA}$ at 50 per cent throughput, cf. $\simeq 1200\,\rm{\AA}$). Thus the presence of plages which could affect $B$-band photometry may not be observed in $\gwfc$-band observations. Furthermore, although the two loci are indistinguishable in the $M_{g_{_{\rm{WFC}}}}, (g-i)_{_{\rm{WFC,\circ}}}$ (when compared to the $V, B-V$) CMD, this could still be consistent with the presence of star spots as the observed effect depends entirely on the adopted spot-to-star contrast \citep{Jackson14}. The lack of significant difference between the two loci at lower masses in the $M_{g_{_{\rm{WFC}}}}, (g_{_{\rm{WFC}}}-K_{\rm{s}})_{\circ}$ CMD disfavours the idea that pre-MS stars are simply inflated by magnetic activity (the active stars would appear much redder), but is consistent with inflation due to the presence of star spots (see e.g. \citealp{Jackson14}).

\subsection{Effects of metallicity on semi-empirical BC-$T_{\rm{eff}}$ relations}
\label{effects_metallicity_semi-empirical_BC-Teff_relations}

The stellar properties predicted by evolutionary models strongly depend on the opacity -- and hence metallicity -- of the stellar interior as this will dictate how the convective energy transfer propagates through the stellar interior. For example, as the metallicity increases so does the opacity, and therefore for an isochrone of a given age, the position in the Hertzsprung-Russell (H-R) diagram will be fainter and cooler when compared to a similarly aged isochrone with a lower metallicity (see e.g. \citealp{Tognelli11}). Not only is the luminosity of an isochrone strongly dependent upon metallicity, but in the CMD the degree of sensitivity further depends on the combination of magnitude and colour index adopted. At low temperatures ($T_{\rm{eff}} \lesssim 3500\,\rm{K}$) molecular species start to become the dominant source of opacity in stellar atmospheres, resulting in noticeable modification to the emergent spectrum. Hence, given the main sources of opacity in this $T_{\rm{eff}}$ regime (e.g. TiO and VO in the optical, and H$_{2}$O in the IR) it is clear that variations in the metallicity could result in appreciable differences to both the stellar spectra and atmospheric structure, and ultimately the observed colours.

The metallicity of Praesepe (in comparison to that of the Pleiades) is approximately 20 per cent supersolar. Unfortunately, the BCAH98 $\alpha=1.9$ and the updated Pisa models are not available in supersolar compositions.
Given that we are limited to solar metallicity interiors, we require an estimate of the effect that using these interiors would have on the semi-empirical BC-$T_{\rm{eff}}$ relations. There are two possible effects which we must consider. First, given this 20 per cent difference between the metallicities of the Pleiades and Praesepe, how much of a difference does this make to the interior models and what systematic errors could we introduce into the semi-empirical model isochrones? Secondly, is the physical process which is responsible for the $\Delta$BCs (when comparing the models and the data in CMDs) strongly dependent on metallicity?

\subsubsection{Effects of metallicity on stellar interior models}
\label{effects_of_metallicity_stellar_interior_models}

\begin{figure}
\centering
\includegraphics[width=0.8\columnwidth]{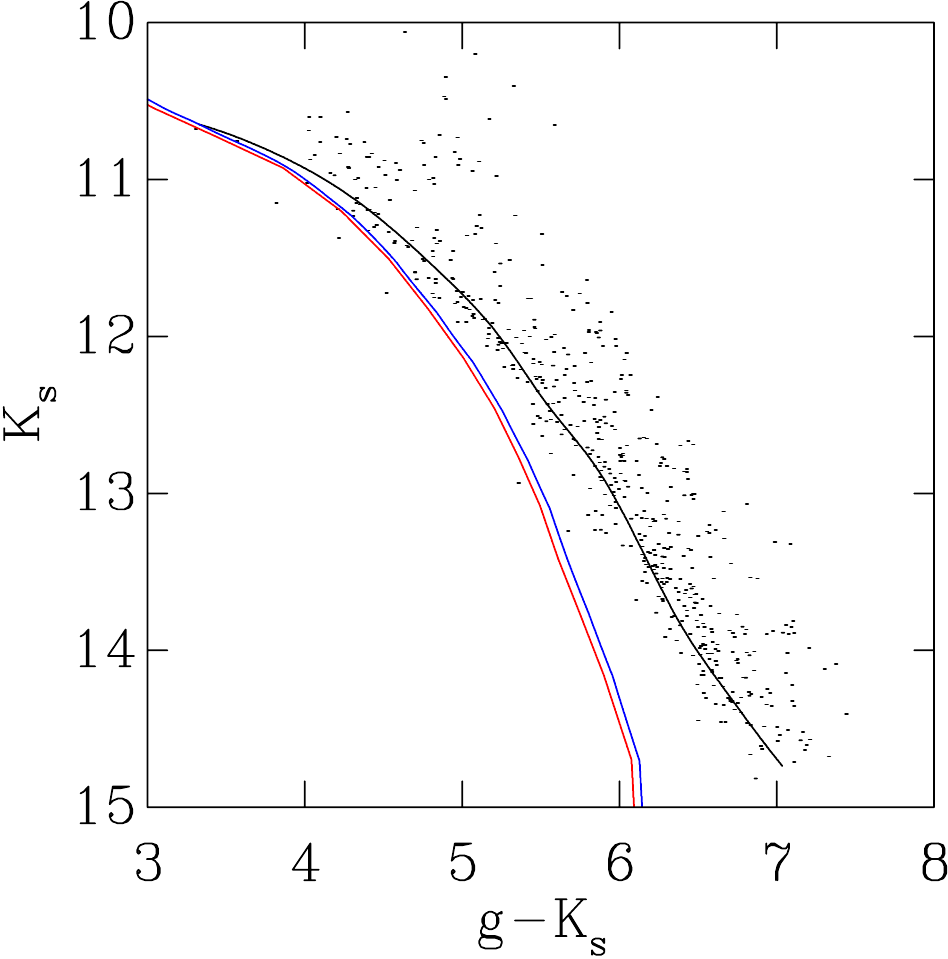}
\caption[]{The effects of different metallicity interior models in the $K_{\rm{s}}, g-K_{\rm{s}}$ CMD of Praesepe members. Overlaid are two $665\,\rm{Myr}$ theoretical (not semi-empirical) DCJ08 model isochrones adopting a distance modulus $dm=6.32\,\rm{mag}$ and a reddening equivalent to $E(B-V)=0.027\,\rm{mag}$. Both model isochrones have been transformed into CMD space using supersolar ($1.2Z_{\odot}$) atmospheric models, however the interior models are solar metallicity ($Z_{\odot}$; \emph{red}) and supersolar ($1.2Z_{\odot}$; \emph{blue}). The black line represents the spline fit to the Praesepe single-star sequence.}
\label{fig:metal_diff}
\end{figure}

Higher metallicity DCJ08 models are available and this therefore allows us to test the effects of metallicity variations on the derived semi-empirical BC-$T_{\rm{eff}}$ relations. Fig.~\ref{fig:metal_diff} shows the Praesepe single-star sequence (as given in Table~\ref{tab:praesepe_sdss}) overlaid on the Praesepe members as defined by \cite{Kraus07} in the $K_{\rm{s}}, g-K_{\rm{s}}$ CMD. We used the positions from \cite{Kraus07} to extract 2MASS photometry for each source from the 2MASS Point Source Catalogue \citep{Cutri03}. Also plotted in the CMD are two $665\,\rm{Myr}$ DCJ08 model isochrones; a solar interior model transformed using appropriately supersolar ($1.2Z_{\odot}$) atmospheric models (\emph{red}) and a supersolar ($1.2Z_{\odot}$) interior transformed using the same supersolar atmospheric models (\emph{blue}). Given that we are unable to derive $\Delta$BCs for fiducial sequences of different metallicity in the SDSS system, the DCJ08 model isochrones shown in Fig.~\ref{fig:metal_diff} are theoretical i.e. not semi-empirical models.

We can quantify the effect that using solar metallicity interior models would have on the derived BC-$T_{\rm{eff}}$ relations by calculating the difference in the colour index at a fixed $K_{\rm{s}}$-band magnitude. In the $T_{\rm{eff}}$ range where we calculate empirical BCs using observed stellar colours, we calculate that the maximum difference between the colour indices is $[g-K_{\rm{s}}]_{1.2Z_{\odot}}-[g-K_{\rm{s}}]_{Z_{\odot}} \simeq 0.05\,\rm{mag}$. In absolute terms, such a difference may appear unacceptably large, however recall that we are recalibrating the model isochrones to correct for a discrepancy of $\sim 0.8\,\rm{mag}$ in $g-K_{\rm{s}}$ at the lowest $T_{\rm{eff}}$ (see Fig.~\ref{fig:metal_diff}).

Investigating whether this difference is wavelength dependent, we quantified the difference in other colour indices and found it to be $\leq 0.05\,\rm{mag}$. This then implies that despite adopting solar metallicity interior models to create semi-empirical BC-$T_{\rm{eff}}$ relations, the resultant semi-empirical model isochrones would have a residual uncertainty of $\lesssim 0.05\,\rm{mag}$ in the magnitude and $\lesssim 0.07\,\rm{mag}$ in the colour.

We provide only solar composition semi-empirical SDSS pre-MS model isochrones through the internet server. In the case of the DCJ08 models, where supersolar interior models are available, we calculate the $\Delta$BCs as described in Section~\ref{semi-empirical_bolometric_corrections} using supersolar (1.2$Z_{\odot}$) interior and atmospheric models and apply these to the theoretical solar composition BC grid. Thus, for the semi-empirical DCJ08 model isochrones we expect the uncertainties in the colours and magnitudes to be smaller than the values quoted above. For both the BCAH98 $\alpha=1.9$ and updated Pisa models, however, we can only calculate the $\Delta$BCs by combining solar interior and supersolar atmospheric models.


\subsubsection{Effects of metallicity on $\Delta$BC}
\label{effects_of_metallicity_deltabc}

Given that we have INT-WFC observations of both the Pleiades and Praesepe, a simple test to investigate whether the physical process responsible for the $\Delta$BCs is strongly dependent upon the metallicity is to use the semi-empirical BC-$T_{\rm{eff}}$ relation we have derived using the Pleiades (solar composition) to transform a solar metallicity interior model at an age of $665\,\rm{Myr}$ and overlay it on the Praesepe photometric data.

\begin{figure*}
\centering
\includegraphics[width=\textwidth]{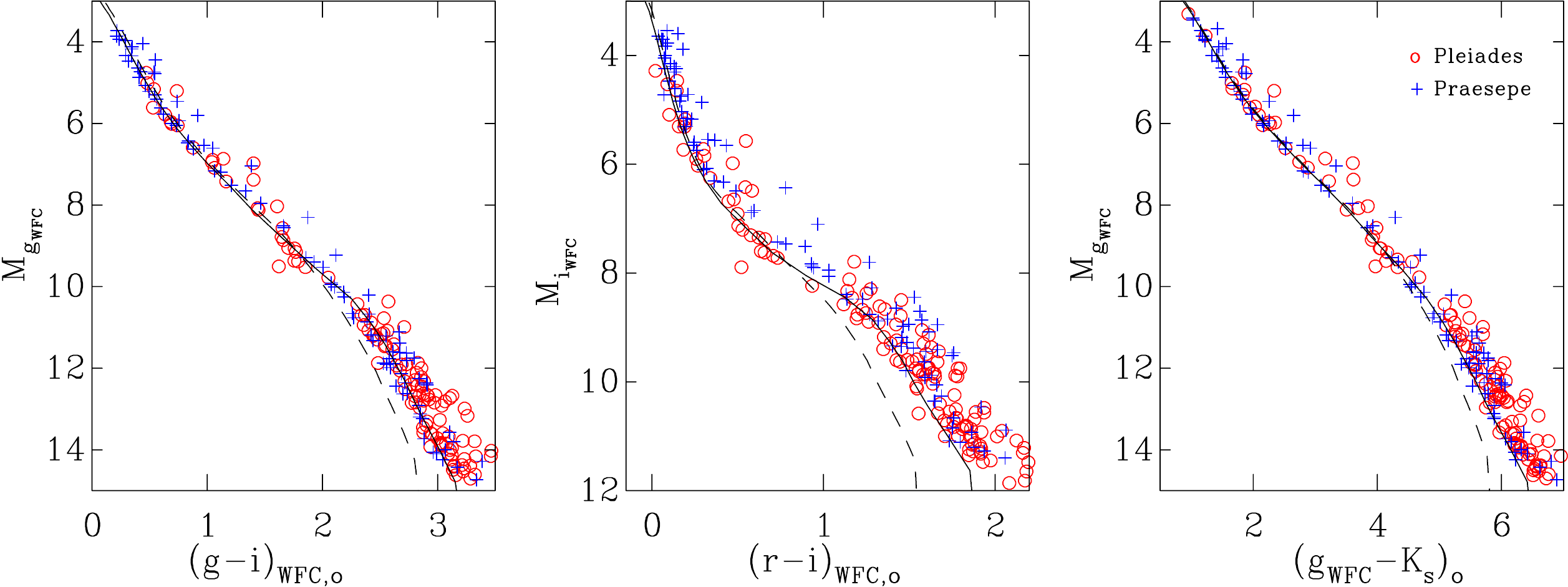}
\caption[]{Same as Fig.~\ref{fig:pleiades_praesepe} but demonstrating that $\Delta$BC is not strongly dependent upon metallicity. The continuous line in each panel is a $665\,\rm{Myr}$ solar composition semi-empirical DCJ08 model isochrone (transformed using the semi-empirical BC-$T_{\rm{eff}}$ relation derived using the Pleiades). The dashed line is a theoretical (not semi-empirical) supersolar $(1.2Z_{\odot})$ metallicity DCJ08 model isochrone. The ``turn-over" in both isochrones ($T_{\rm{eff}} \simeq 3200\,\rm{K}$) is an artefact of the DCJ08 interior models at older ages which is not present at the age of the Pleiades and hence our log$\,g$-dependent $\Delta$BCs are unable to account for this.}
\label{fig:metal_delta_bc}
\end{figure*}

Fig.~\ref{fig:metal_delta_bc} is an adaptation of Fig.~\ref{fig:pleiades_praesepe} with two $665\,\rm{Myr}$ DCJ08 model isochrones overlaid. The continuous line is a solar composition semi-empirical model isochrone (transformed using the semi-empirical BC-$T_{\rm{eff}}$ relation derived using the Pleiades), whereas the dashed line is a theoretical (not semi-empirical) supersolar ($1.2Z_{\odot}$) metallicity model isochrone. The reason we combine both photometric datatsets is simply to increase the number of stars in our sequence. Due to the small areal coverage of our observations in Praesepe there is a paucity of stars in certain regions along the sequence, and so by combining this with our Pleiades data we can better assess whether the isochrones represent a good match to the observed sequence.

It is clear from Fig.~\ref{fig:metal_delta_bc} that the solar composition DCJ08 model isochrone is a good fit to the observed shape of the Praesepe sequence in all colours. Thus, we can be reasonably assured that despite composition discrepancies (again at least at the 20 per cent level) between the photometric data and the model isochrones adopted (see Section~\ref{effects_of_metallicity_stellar_interior_models}), this should not have a significant effect on the age derived from using such isochrones in CMDs.

Note that in Fig.~\ref{fig:metal_delta_bc} there is an obvious mismatch between the data and the models at low $T_{\rm{eff}}$. At the age of Praesepe and temperatures below $\simeq 3200\,\rm{K}$, the DCJ08 interior models become almost vertical in the H-R diagram (see also Fig.~\ref{fig:metal_diff}). This ``turn-over" in the models is not observed at younger ages e.g. the Pleiades, and therefore our method of creating semi-empirical BC-$T_{\rm{eff}}$ relations based on the discrepancy between the models and the data will not account for this effect at older ages. Hence we urge the user to exercise caution if using the semi-empirical DCJ08 model isochrones at old ages and at low $T_{\rm{eff}}$. This effect is not observed in either the BCAH98 $\alpha=1.9$ or updated Pisa models.

\section{Pre-MS isochrone server}
\label{internet_server}

By construction the semi-empirical pre-MS model isochrones presented here will deliver reliable ages between 600\,Myr (the age of Praesepe) and 100\,Myr (roughly the age of the Pleiades). The question is how well they will work for younger ages, which amounts to testing our assumption that the theoretical relationship for the change in BC with log$\,g$, can be applied to the empirical corrections derived from the Pleiades and Praesepe. The next youngest test we have is NGC\,1960, which has a LDB age of $22\pm4\,\rm{Myr}$ \citep{Jeffries13}, and an upper MS age of $26.3^{+3.2}_{-5.2}\,\rm{Myr}$ (Paper~II). The isochrones presented here based on the BCAH98 $\alpha=1.9$ or DCJ08 interior models yield an age of $19-21\,\rm{Myr}$ (Paper~II) in good agreement with the other two age diagnostics. Unfortunately, there are no LDB ages for younger clusters, and thus the only remaining test is the upper MS ages, which typically have large age uncertainties. Of these, however, $\lambda$ Ori has a rather precise upper MS age of $9-11\,\rm{Myr}$ (Paper~II), which is in good agreement with the ages from the present semi-empirical models isochrones of $10.0-11.0\,\rm{Myr}$ (BCAH98 $\alpha=1.9$ interiors) or $7.6-8.7\,\rm{Myr}$ (DCJ08 interiors). There are five other clusters with ages of between 5 and 15\,Myr where the uncertainties in the upper MS age are of order a factor two, but within these uncertainties the upper MS ages and those for the isochrones in this work agree. Only at around 3\,Myr is there significant evidence for a difference between the pre-MS and upper MS ages (Paper~II). Thus the semi-empirical pre-MS model isochrones presented here yield ages in agreement with other methods over the two decades of age from 6 to 600\,Myr.

We reiterate the three underlying assumptions that we have adopted in the construction of these semi-empirical model isochrones. \emph{Assumption 1: the theoretical models fit the $K_{\rm{s}}$-band flux well and therefore we can use this to determine the $T_{\rm{eff}}$ at points along our fiducial locus and derive the necessary corrections. Assumption 2: the empirical corrections in a given bandpass have the same log$\,g$ dependence as the theoretical BCs. Assumption 3: all clusters are assumed to behave like the Pleiades such that there are no intracluster effects due to, for example, binary fractions and mass ratios, intrinsic age spread, rotation distribution, and activity-related effects on the observed colours.} It is important that these assumptions are well understood by the reader/user given that our primary aim in making these models available is to allow other users to calculate cluster ages from the pre-MS members which are on a consistent scale with those we have derived (see Paper~II). These models are also tailored for identifying possible new members and determining mass functions for stellar populations from their positions in the CMD. Given the methodology of creating these model isochrones, we would advise users \emph{not} to use these models to compare against new photometric observations and then critique either the interior or atmospheric models due to a mismatch between the models and the data.

Our semi-empirical pre-MS model isochrones can be found at \url{http://www.astro.ex.ac.uk/people/timn/isochrones/}. The server itself is designed to be self-explanatory and guide the user through the various choices e.g. pre-MS interior model, mass and age range, and photometric system, etc., however we would like to provide a little more detail on two points, namely the treatment of the reddening and extinction, and the output of the model isochrones.

\subsection{Reddening and extinction}
\label{reddening_extinction}

Using the simple formalism we have adopted for deriving the BCs from atmospheric models (see
Section~\ref{atmospheric_models}), it is 
straightforward to then include the effects of both reddening and extinction when creating a model isochrone. As discussed in Paper~II (see also \citealp*{Bessell98}) the reddening and extinction applied to each mass point in a given isochrone is dependent upon the $T_{\rm{eff}}$ of the star i.e. for a given amount of intervening material, a larger value of, for example, $E(B-V)$ will be measured from the higher $T_{\rm{eff}}$ stars than from the lower $T_{\rm{eff}}$ objects. This therefore leads to differential reddening along the model isochrone and we can account for such effects in the isochrone server by following the  method described in Paper~II, whereby we construct extinction grids based on the
atmospheric models, photometric system bandpasses and a description of
the interstellar extinction law.

The atmospheric models were reddened adopting the interstellar
extinction law of \cite{Fitzpatrick99} with $R_{V}=3.1$, and folded
through the various
system responses (see Table~\ref{tab:photometric_bandpasses} for a
list). The models were reddened in steps of 0.5 from
$E(B-V)_{\rm{nom}}=0.0$ to $2.0\,\rm{mag}$ [where $E(B-V)_{\rm{nom}}$
represents the amount of interstellar material between an observer and
the object], with the grids comprising
extinction in the various bandpasses as a function of $T_{\rm{eff}}$
and log$\,g$. Thus to redden an isochrone, the user can specify a specific
$E(B-V)$ value and the server will
automatically calculate the corresponding $E(B-V)_{\rm{nom}}$, and
subsequently use this to interpolate within the extinction grid for
the extinction and reddening for a star of given $T_{\rm{eff}}$ and
log$\,g$ in the model isochrone. This process is then repeated along
the full mass range of the isochrone. 


\subsection{Model isochrone output}
\label{model_isochrone_output}

The standard model isochrone output comprises a single-star sequence
consisting of $10\,000$ points equally sampling a given mass
range. The user, however, may also wish to
include the effects of binarity, essentially modifying the output from a linear curve in CMD space to a two-dimensional distribution which can then be used to perform statistically robust fitting to photometric datasets
(e.g. \citealp{Cargile10}; \citealp*{DaRio10b}).

We follow the formalism as mentioned in Section~\ref{fiducial_loci_parameters} (see Paper~II for a more complete discussion), whereby we create a two-dimensional
distribution using a Monte Carlo method to simulate $10^{6}$ stars
over a specific mass range for a given mass function and binary fraction. The mass function we adopt is the canonical broken power law IMF of \cite{Dabringhausen08} discussed in Section~\ref{praesepe_parameters}. The user can then choose a given binary fraction (between 0.0 and 1.0), and if a generated star happens to be
a binary, the companion mass is assigned as described in
Section~\ref{fiducial_loci_parameters}.
Note that due to lower $T_{\rm{eff}}$ limits as a result of deriving the empirical BCs (or alternatively mass limits in the interior models) some assigned secondary stars may lie below this lower
limit. In such cases, these secondary stars are assumed to make a negligible
contribution to the system light, which is equivalent to placing the
binary on the single-star sequence. This limitation can lead to a wedge
of zero probability between the single- and binary-star sequences at
low masses.

\section{Conclusions}
\label{conclusions}

We present an isochrone server to make sets of semi-empirical pre-MS isochrones in the Johnson-Cousins, INT-WFC, 2MASS, SDSS, and IPHAS/UVEX photometric systems publicly available. The stages we have gone through to achieve this are as follows:

\begin{itemize}
  \item We created fiducial loci in CMD space using photometric data of young clusters with known age, distance and reddening. For the Johnson-Cousins, INT-WFC, 2MASS, and IPHAS/UVEX systems we adopted the Pleiades, whereas for the SDSS system we used Praesepe.
  \item Using the $K_{\rm{s}}$-band magnitude to define the $T_{\rm{eff}}$ scale along the fiducial loci, we quantified the discrepancy between the observed sequence and theoretical pre-MS model isochrones as a function of $T_{\rm{eff}}$ in individual photometric bandpasses.
  \item Semi-empirical pre-MS model isochrones can then be created using existing stellar interior models in conjunction with the recalibrated BC-$T_{\rm{eff}}$ relations (assuming a model dependency for log$\,g$). 
  \item These new semi-empirical models are made available via the Cluster Collaboration website
    \url{http://www.astro.ex.ac.uk/people/timn/isochrones/} and the
    user is able to choose the following inputs: interior model,
    mass range, age (or age range), and photometric system, as well as
    include the effects of interstellar reddening and binarity. In the
    future, we expect to make additional interior models and
    photometric systems available through this server.
  \item We have investigated the effects of both increased magnetic activity and variations in chemical compositions between using the Pleiades and Praesepe to define our fiducial loci. \emph{We find no evidence, through comparison in CMD space, to suggest that one locus is systematically offset with respect to the other}. Furthermore, we note that the colours of the semi-empirical pre-MS model isochrones in the SDSS photometric system have residual uncertainties of $\simeq 0.07\,\rm{mag}$ as a result of the metallicity discrepancy between the Praesepe photometric data and the stellar interior models.
  \item We have derived new cluster parameters for both the Pleiades and Praesepe, which are on a consistent scale with the MS parameters given in Paper~II. Not allowing for systematic uncertainties, we determine ages of $135^{+20}_{-11}$ and $665^{+14}_{-7}\,\rm{Myr}$ as well as distances of $132 \pm 2$ and $184 \pm 2\,\rm{pc}$ for the Pleiades and Praesepe respectively.
\end{itemize}

\section*{Acknowledgements}

JMR is funded by a UK Science and Technology Facilities Council
(STFC) studentship. EEM acknowledges support from the National
Science Foundation (NSF) Award AST-1008908. The
authors would like to thank Emanuele Tognelli for the updated set of
Pisa models and John Stauffer for sharing his catalogue of
Kron photometric measurements of Pleiades members. The authors
would also like to thank the referee for comments which have vastly
improved the clarity of the manuscript.

This research has made use of data obtained at the
Isaac Newton Telescope which is operated on the island of La
Palma by the Isaac Newton Group (ING) in the Spanish
Observatorio del Roque de los
Muchachos of the Institutio de Astrofisica de Canarias.
This research has made use of archival data products from the
Two-Micron All-Sky Survey (2MASS), which is a joint project of the
University of Massachusetts and the Infrared Processing and Analysis
Center, funded by the National Aeronautics and Space Administration (NASA)
and the National Science Foundation.

This research has made use of public data from the SDSS.
Funding for the SDSS was provided by the Alfred P. Sloan Foundation,
the Participating Institutions, the National Science Foundation, the U.S. Department
of Energy, the National Aeronautics and Space Administration, the Japanese
Monbukagakusho, the Max Planck Society, and the Higher Education Funding
Council for England. The SDSS was managed by the Astrophysical Research
Consortium for the Participating Institutions.

\bibliographystyle{mn3e}
\bibliography{references}

\label{lastpage}

\end{document}